\documentclass[iop,apj]{emulateapj}
\usepackage{natbib,amsmath,multirow}
\DeclareMathOperator{\arctanh}{arctanh}
\newcommand{\mf}{{\cal M}}

\newcommand{\mcmax}{{M_{c, max}}}

\begin{document}

\title{A Posteriori Transit Probabilities}
\author{Daniel J. Stevens and B. Scott Gaudi}
\affil{Department of Astronomy, The Ohio State University, 140 W. 18th
Ave., Columbus, OH, USA 43210} 

\begin{abstract}
Given the radial velocity (RV) detection of an unseen companion, it is
often of interest to estimate the probability that the companion also
transits the primary star.  Typically, one assumes a uniform
distribution for the cosine of the inclination angle $i$ of the
companion's orbit.  This yields the familiar estimate for the {\it
prior} transit probability of $\sim R_*/a$, given the primary radius
$R_*$ and orbital semimajor axis $a$, and assuming small companions
and a circular orbit.  However, the {\it posterior} transit
probability depends not only on the prior probability distribution of
$i$ but also on the prior probability distribution of the companion
mass $M_c$, given a measurement of the product of the two (the minimum mass $M_c \sin i$)
from an RV signal.  In general, the posterior can be larger or smaller than
the prior transit probability. We derive analytic expressions for the
posterior transit probability assuming a power-law form for the
distribution of true masses, $d\Gamma/dM_c \propto M_c^\alpha$, for
integer values $-3 \le \alpha \le 3$. We show that for low transit
probabilities, these probabilities reduce to a constant multiplicative
factor $f_\alpha$ of the corresponding prior transit probability,
where $f_\alpha$ in general depends on $\alpha$ and an assumed upper
limit on the true mass.  The prior and posterior probabilities are
equal for $\alpha = -1$.  The posterior transit probability is
$\sim 1.5$ times larger than the prior for $\alpha = -3$ and is
$\sim 4/\pi$ times larger for $\alpha = -2$, but is less than the
prior for $\alpha \geq 0$, and can be arbitrarily small for $\alpha >
1$. We also calculate the posterior transit probability in different
mass regimes for two physically-motivated mass distributions of
companions around Sun-like stars. We find that for Jupiter-mass
planets, the posterior transit probability is roughly equal to the
prior probability, whereas the posterior is likely higher for
Super-Earths and Neptunes $(10 M_\oplus - 30 M_\oplus)$ and
Super-Jupiters $(3 \textrm{M}_{\rm Jup}-10 \textrm{M}_{\rm Jup})$,
owing to the predicted steep rise in the mass function toward smaller masses in
these regimes.  We therefore suggest that companions with minimum
masses in these regimes might be better-than-expected targets for
transit follow-up, and we identify promising targets from
RV-detected planets in the literature.  Finally, we consider the
uncertainty in the transit probability arising from uncertainties in
the input parameters, and the effect of ignoring the dependence of 
the transit probability on the true semimajor axis on $i$.
\end{abstract}

\section{Introduction}\label{sec:intro}

Transiting planets have become the primary resource for characterizing
the detailed properties of exoplanets.  When complemented by RV
measurements of the planet's orbital eccentricity, period, argument of
periastron, and velocity semiamplitude, as well as by measurements of
the stellar host's mass and radius, a photometric transit allows for
the measurement of the mass, radius, density, and surface gravity of
the planet.  With these basic parameters in hand, a transiting planet
system is then amenable to a profusion of follow-up
observations, which can then yield an impressive array of physical
properties of the planet and star (see \citealt{Winn2011-2} for a
review).

Transiting planets are currently discovered via two different methods.
First, planets that are initially discovered by RV measurements are
monitored photometrically during the predicted time of inferior
conjunction, in order to detect the small fraction that transit.  In
fact, the first identified transiting planet HD 209458b was discovered
in this manner \citep{Charbonneau2000,Henry2000}.  Second, photometric
transit surveys synoptically monitor large numbers of stars to search
for eclipse signals consistent with planetary-sized companions.  These
candidate transiting planets are then subjected to a battery of
follow-up observations in order to eliminate false positives and
ultimately confirm the planet by measuring its RV signal and, in doing
so, its mass.

Because it is relatively easy and inexpensive to monitor large numbers
of stars photometrically with the precision needed to detect planetary
transit signals, photometric surveys for transiting planets can
readily overcome the intrinsic rarity of transiting systems and
efficiently identify large samples of candidates.  Indeed, the vast
majority of confirmed transiting planets were discovered in dedicated
ground-based transit surveys (e.g.,
\citealt{udalski2002,konacki2003,alonso2004,mccullough2006,bakos2007,
cameron2007,weldrake2008,alsubai2011, siverd2012}).  In addition,
Kepler has discovered over 2,700 transiting planet candidates
\citep{batalha2013,burke2013} --- an amount greater than that of all
previously-known exoplanets, although most of these lack precision RV
confirmation and so lack precision mass estimates.  However, the
cadence, photometric precision, and total number of observations
needed to achieve robust detections of transit signals typically
requires monitoring relatively small fractions of the sky at a time;
as a result, photometric transit surveys typically identify transiting
planets orbiting relatively faint ($V\ga 10$) stars.

On the other hand, because of the relatively demanding observational
requirements for precision RV measurements --- in particular,
high-resolution spectra, broad wavelength coverage, and a high
signal-to-noise ratio --- RV surveys must, in general, target bright
stars in series with relatively large aperture telescopes.  Because of
these constraints, only a few thousands of stars have been monitored
with precision RV.  When combined with the rarity of planetary
companions in general and transiting planets in particular, only eight
of the planets originally detected via RV have been subsequently shown
to transit \citep{Barbieri2007, Winn2011, Demory2011,Charbonneau2000,
Henry2000,
Sato2005,Bonfils2012,Bouchy2005,Moutou2009,fossey2009,Gillon2007}.
Nevertheless, because they all orbit bright stars, these systems are
the most amenable to follow-up programs and so are extraordinarily
valuable.  Indeed, these systems are some of best-characterized
planets outside our own solar system.

Fortunately, it is likely that the both the sample size and diversity
of transiting planets originally discovered by precise RV surveys will
expand considerably in the future via several avenues.  Amongst the
currently-known RV planets, it is expected that there exist a handful
of long-period transiting planets that have not yet been
photometrically identified.  The Transit Ephemerides Refinement and
Monitoring Survey \citep{Kane2010} aims to identify these systems by
first refining the orbits (and the estimated time of conjunction) of
the most promising systems (i.e. those with the highest transit
probabilities) and then following these up photometrically
\citep{Seagroves2003}.  Even more promising, samples of RV-detected
planets have been and will continue growing significantly.  The
measurement precision of optical RV surveys has steadily improved,
allowing for the detection of first Neptune-mass
\citep{mcarthur2004,Butler2004,lovis2006}, then Super-Earth mass
\citep{rivera2005,udry2007,mayor2009}, and most recently Earth-mass
\citep{dumusque2012} companions, all of which are intrinsically more
numerous than gas giant planets \citep{howard2010,mayor2011}.  In the
near future, the development of dedicated near-IR precision RV
instruments (e.g.,
\citealt{Bean2010,Plavchan2013,Rayner2007,mahadevan2012}) will enable
surveys of large, previously-inaccessible samples of low-mass stars,
and will also have enhanced sensitivity to low-mass companions.
Finally, massive parallel RV surveys, such as the Multi-object APO
Radial Velocity Exoplanet Large-area Survey (MARVELS)
\citep{ge2008,eisenstein2011}, may allow for surveys of giant planets
orbiting somewhat fainter but considerably more numerous stars.

It is likely that substantial photometric follow-up resources will be
necessary to identify the transiting systems from these anticipated
large samples of RV-detected exoplanets.  For example, the
CHaracterizing ExoPlanets Satellite (CHEOPS), planned for launch in
2017, is a satellite dedicated just to this purpose \citep{Feldt2007}.
It is important that these necessarily-limited photometric follow-up
resources be allocated optimally to maximize the yield of transiting
planets.

The most important input to optimizing follow-up efforts is an
estimate of the transit probability $P_{tr}$.  The transit probability
is defined as the probability, given both the properties of the
companion that are inferred from RV measurements and the properties of
the host star that are inferred from auxilliary measurements, that the
orbit of the companion is inclined such that it passes in front of the
host star from our perspective at some point during its orbit.  The
simplest and most commonly used estimate of the transit probability is
\citep{Borucki1984,Sackett1999},
\begin{equation}
\label{eq:P0}
P_{tr} = \frac{R_*}{a},
\end{equation}
where $R_*$ is the radius of the host star and $a$ is the semimajor
axis of the companion's orbit.  This estimate makes a number of
assumptions that are widely appreciated, including: a circular orbit
for the companion, a companion radius $R_c$ that is much smaller than
the stellar radius, and a uniform distribution for the cosine of the
inclination angle $i$ of the orbit.  The effect of the finite size of
the planetary companion can be easily included by, e.g., replacing
$R_* \rightarrow R_*+R_c$ for grazing transits.  The effect of
eccentric orbits on the transit probability has been explored in a
number of papers \citep{Seagroves2003, Barnes2007, Burke2008,
Kane2008}.  Eccentricity can boost the transit probability relative to
the naive estimate in Equation \ref{eq:P0}, particularly for some
favorable geometries.  HD 17156 b, with its eccentricity of 0.67
\citep{fischer2007,Barbieri2007}, and HD 80606 b, with an eccentricity
of 0.927 \citep{Naef2001,Moutou2009,fossey2009}, are dramatic examples
of these effects.  \citet{BeattySeager2010} found that a prior
constraint on the inclination of the host star can also boost the
transit probability (see also \citealt{Sackett1999,Watson2010}),
assuming the axis of the companion orbit is aligned with the spin axis
of the star.  Finally, \citet{Kane2008} showed that constraints on
secondary eclipses can also affect the transit probability.

However, there are some additional assumptions inherent in Equation
\ref{eq:P0} that are less widely appreciated and that can also affect
the transit probability; we explore several of these here.  In
particular, the {\it posterior} probability distribution of the orbit
inclination angle $i$, given the detection of an RV companion, depends
not only on the prior distribution of $i$ but also on the prior
distribution of the true mass of the companion $M_c$.  While the prior
distribution of $i$ is well-known (and is simply uniform in $\cos i$),
the prior distribution of $M_c$ is generally not well-constrained ---
at least, not in the regimes of interest for exoplanet surveys.
Although the effect of the prior distribution of $M_c$ on the transit
probability has been noted and estimated previously for a few specific
cases (e.g., \citealt{Wisniewski2012}), to the best of our knowledge
it has not been explored in any detail.  Previous authors have noted
the dependence of the posterior probability distribution of $i$ (and
thus $M_c$) on the prior distributions of $i$ and $M_c$ (e.g.,
\citealt{Ho&Turner2011,Lopez&Jenkins2012}).  The fact that the transit
probability also depends on both these priors follows trivially from
this result.  In addition, here we explore two additional effects that
we also believe have not previously been discussed in detail.  First,
we estimate the effects of uncertainties on the RV observables and
host star properties on the estimated transit probability.  Second, we
note that the semimajor axis in Equation \ref{eq:P0} is properly the
true semimajor axis (rather than the minimum semimajor axis), which is
not known from RV measurements alone, and indeed depends on the
orbital inclination for a given set of observables.

The plan for this paper is as follows.  In Section \ref{sec:setup}, we
review the observables for RV detected companions, and reiterate how
these can be used to derive the familiar \emph{a priori} transit
probability from RV data.  In Section \ref{sec:derivation}, we derive
the \emph{a posteriori} transit probability for an observed minimum
mass $M_0$ and true mass distribution $\text{d}N/\text{d}M_c$. In
Section \ref{sec:powerlaw}, we explore the posterior transit
probability for power-law mass distributions, discuss issues of
convergence for certain power-law distributions, and show that the
posterior transit probability can be expressed as a simply scaling of
the prior probability for low transit probabilities. We calculate
posterior transit probabilities for different regimes of two model
companion mass distributions in Section \ref{sec:bd}. In Section
\ref{sec:RVs}, we compare the number of known transiting RV-discovered
transiting planets with the number expected based on the naive prior
transit probability.  We also compare the posterior and prior transit
probabilities for known RV planets, and identify planets whose
posterior transit probabilities are both high and substantially higher
than the corresponding naive prior probabilities.  In Section
\ref{sec:issues}, we discuss the effect of propagating parameter
uncertainties on the estimate of the transit probability, as well as
the effect of assuming the minimum semimajor axis to estimate the
transit probability.  We summarize our results in Section
\ref{sec:conclusion}.

\section{Problem Set-up}\label{sec:setup}
Consider the RV detection of a faint companion,
such as a single-lined spectroscopic binary.  A well-sampled, high
signal-to-noise ratio RV detection yields estimates of the orbital
period $T$, semi-amplitude $K$, eccentricity $e$, the argument of
periastron of the host star $\omega$, and the time at some reference
point in the orbit (e.g. the periastron).  The semi-amplitude $K$ is
related to $T$, $e$, the host star mass $M_*$, the companion mass
$M_c$, and the orbital inclination $i$ via
\begin{equation}
\label{eqn:Kdef}
K = \left( \frac{2\pi G}{T} \right)^{1/3}\frac{M_c \sin i }{(M_* + M_c)^{2/3}}(1-e^2)^{-1/2}.
\end{equation}

As is well known, the only model-independent physical quantity one can
infer about a single-lined spectroscopic binary purely from direct
observables is the mass function $\mf$, which is defined as
\begin{equation}
\mf \equiv \frac{(M_c \sin i)^3}{(M_* + M_c)^2} = \frac{K^3 T}{2\pi G}(1-e^2)^{3/2}.
\label{eqn:massfunc}
\end{equation}
Generally, one can also estimate the mass (and radius $R_*$) of the
host star through a variety of methods.  With an estimate of $M_*$ one
can use the measured value of $\mf$ to determine $M_c$ as a function
of $i$ by solving the resulting cubic equation for $M_c$.  One can
then also estimate the semimajor axis of the orbit as a function of
$i$,
\begin{equation}
a = \left(\frac{G[M_*+M_c(i)]}{4\pi^2}\right)^{1/3} T^{2/3}
\label{eqn:semi}
\end{equation}

If one has independent reason to believe that $M_c \ll M_*$, then the
``minimum mass'' of the companion can be estimated directly by,
\begin{eqnarray}
M_0 &\equiv& M_c\sin i\nonumber\\
&=& K \left(\frac{T}{2\pi G}\right)^{1/3}(1-e^2)^{1/2}(M_*+M_c)^{2/3}\nonumber\\
&\simeq& K \left(\frac{T}{2\pi G}\right)^{1/3}(1-e^2)^{1/2}M_*^{2/3}.
\label{eqn:minimummass}
\end{eqnarray}

An excellent approximation for the condition that the companion 
transits the host is that its projected separation from the center of
the host star at the time of inferior conjunction is less than or
equal to the sum of $R_*$ and the radius of the
companion $R_c$ \citep{Kipping2008, Winn2011-2}.  The orbital separation of
the companion from the host at the time of inferior conjunction is
\begin{equation}
r_c = \frac{a(1-e^2)}{1+e\sin \omega},
\label{eqn:rc}
\end{equation}
leading to the condition for a transit,
\begin{equation}
r_c \cos i \le R_* + R_c.
\label{eqn:condtrans}
\end{equation}
Or, in terms of a limit on the inclination,
\begin{equation}
\label{eq:condition}
\cos i \leq \left(\frac{R_* + R_c}{a}\right)\left(\frac{1 + e\sin\omega}{1-e^2}\right).
\end{equation}

It is straightforward to demonstrate that, for
isotropically-distributed orbit normals, the probability density
distribution of $\cos i$ is uniform.  In the absence of any
information, we expect the orbits of binary systems to be isotropic,
i.e. to have no preference for a given orientation.  Therefore, the
\emph{a priori} probability density distribution of $\cos i$ is
expected to be uniform, and the \emph{a priori} transit probability
is given by
\begin{equation}
\label{eq:priorP}
P_{tr,0} = \left(\frac{R_* + R_c}{a}\right)\left(\frac{1 + e\sin\omega}{1-e^2}\right).
\end{equation}
In the case of a circular orbit and $R_c \ll R_*$, we recover the 
familiar transit probability $P_{tr,0} = R_*/a$. 

However, once one makes a measurement of the companion's minimum
mass\footnote{More precisely, once makes a measurement of the mass
  function.} $M_0$ --- i.e. the product of $M_c$ and $\sin i$ --- the
transit probability then depends not only on the prior probability
density distribution of $i$ but also on the prior probability density
distribution of $M_c$.  That the transit probability depends on the
prior on $M_c$ can be seen intuitively using the following example:
Assume that one had prior knowledge that objects of mass $<M_{min}$
did not exist.  Then, if one detected a companion with minimum mass
$M_0 = 0.1 M_{min}$, one would be certain that the true mass $M_c$ was
at least ten times larger than the minimum mass, and therefore that
$\sin i<0.1$, or $\cos i>0.995$.  Provided that the prior transit
probability $P_{tr,0} < 0.995$, the true \emph{a posteriori} transit
probability would be zero, since one would be certain that object did
not transit.

We derive the general expression for the \emph{a posteriori} transit
probability in the next section.  Before moving on, however, we note
one subtlety that we ignored in the previous discussion.  The
semimajor axis $a$ in equation \ref{eq:priorP} is the true semimajor
axis, which depends on $M_c$ and thus on $i$.  Therefore, in order to
estimate the transit probability, one must either solve simultaneously
for $a$ and the critical inclination for a transit, or use an
approximation to $a$.  We return to this point in Section \ref{subsec:a}.

\section{\emph{A Posteriori} Transit Probability}\label{sec:derivation}
Bayes' Theorem gives the conditional probability density of event A
given event B as 
\begin{equation}
P(A|B) = \frac{P(B|A)P(A)}{P(B)}.
\label{eqn:bayes}
\end{equation}
For our case, $A = \cos
i$ and $B = M_0 $. Then the probability density of $\cos i$
\emph{given} a measurement of $M_0$ is
\begin{equation}
\label{eq:bayes}
P(\cos i|M_0) = \frac{P(M_0|\cos i)P(\cos i)}{P(M_0)},
\end{equation}
where $P(M_0) = \int P(M_0|\cos i)P(\cos i) \textrm{d} \cos i$ and the
integral is over the allowed range of $\cos i$. If we consider some
maximum true mass $\mcmax$, then the integral is evaluated from $\cos
i = 0$ to some minimum inclination angle $i_{min}$ such
that $\cos i_{min} = \sqrt{1 - (\sin i_{min})} = [1-(M_0/\mcmax)^2]^{1/2}$.

The prior probability density of $\cos i$ is uniform; we write this as
\begin{equation}
\label{eq:cosiprior} \frac{\textrm{d}\Gamma}{\textrm{d}\cos i} =
P(\cos i) = \textrm{const}.
\end{equation} 
Here, $\Gamma$ defines the rate or distribution, so
$\textrm{d}\Gamma/\textrm{d}\cos i$ is the prior probability density
with respect to $\cos i$. The probability density of $M_0$ given $\cos
i$, $P(M_0|\cos i)$, is just the prior probability density of $M_p$
times the Jacobian between $M_c$ and $M_0$, i.e.  
\begin{equation} \label{eq:chain} 
P(M_0|\cos i) =
\frac{\textrm{d}\Gamma}{\textrm{d}M_c}\frac{\textrm{d}M_c}{\textrm{d}M_0}
= \frac{\textrm{d}\Gamma}{\textrm{d}M_c}\frac{1}{\sin i},
\end{equation} 
where $\textrm{d}\Gamma/\textrm{d}M_c = P(M_c)$ is the
prior probability density of the true mass $M_c$. Thus,
\begin{equation}
\label{eq:postprob1} P(\cos i|M_0) =
\frac{\frac{\textrm{d}\Gamma}{\textrm{d}M_c}\frac{1}{\sin i}}{\int_0^{\cos i_{min}}\frac{\textrm{d}\Gamma}{\textrm{d}M_c}\frac{1}{\sin i}
\textrm{d} \cos i}.
\end{equation} 
Finally, the probability that the companion transits is
the (cumulative) probability that $\cos i$ is less than the critical
value for a transit, which is
\begin{equation}
\label{eq:tprob1} P_{tr}(\cos i \leq X) =
\frac{\int_0^X\frac{\textrm{d}\Gamma}{\textrm{d}M_c}\frac{1}{\sin i} \textrm{d} \cos i}{\int_0^{\cos
i_{min}}\frac{\textrm{d}\Gamma}{\textrm{d}M_c}\frac{1}{\sin i}
\textrm{d} \cos i}.
\end{equation} 
Here, we have defined the ratio of sum of the radii to
$r_c$, i.e., the prior transit probability, as
\begin{equation} X \equiv \frac{R_*+R_c}{r_c} = \left( \frac{R_* +
R_c}{a} \right) \left(\frac{1 + e\sin\omega}{1-e^2}\right),
\label{eqn:defX}
\end{equation}

\section{Posterior Transit Probabilities for  Power-law Mass Distributions}\label{sec:powerlaw}
\subsection{Functional Forms}\label{sec:exact}
Assume that $d\Gamma/dM_c = CM_c^\alpha$, where C is a normalization
constant. Then $d\Gamma/dM_c = C\left(\frac{M_0}{\sin i}\right)^\alpha
= CM_0^\alpha(1-\cos^2 i)^{-\alpha/2}$ and Equation \eqref{eq:tprob1}
becomes
\begin{equation}
\label{eq:tprob}
P_{tr} = \frac{\int_0^X (1 - \cos^{2}i)^{-\frac{(\alpha+1)}{2}} \textrm{d}\cos i}{\int_0^{\cos i_{min}} (1 - \cos^{2}i)^{-\frac{(\alpha+1)}{2}} \textrm{d}\cos i}.
\end{equation}
It is clear from this expression that, for $\alpha=-1$
(a uniform distribution in $\log M_c$), the posterior transit
probability is equal to the prior transit probability. The integrand in
the denominator is evaluated up to $\cos i_{min}$ to allow for an
upper mass limit on the prior distribution of $M_c$.  This is strictly
necessary for $\alpha \geq 1 $ because this integral diverges.  For
$\alpha < 1$, the integral converges, and so for simplicity we will
take $\cos i_{min}=1$ for these cases.  We discuss the effect of the
choice of $i_{min}$ (or $\mcmax$) on the transit
probability for $\alpha \geq 1$ in Section
\ref{subsubsec:Convergence}. 

Since $X$ and $\cos i_{min}$ are never
greater than one, we can re-write Equation \eqref{eq:tprob} as
\begin{equation}
\label{eq:hyperprob}
P_{tr} = \frac{X[_2F_1(\frac{1}{2},\frac{\alpha + 1}{2};\frac{3}{2};X^2)]}{_2F_1(\frac{1}{2},\frac{\alpha+1}{2};\frac{3}{2};\cos^2 i_{min})},
\end{equation}
where $_2F_1(a,b;c;d)$ is the Gaussian (or ordinary) hypergeometric function:
\begin{equation}
\label{eq:2F1}
_2F_1(a,b;c;d) = \sum\limits_{k=0}^\infty\frac{(a)_k(b)_kd^k}{(c)_kk!},
\end{equation}
where $(x)_n$ is the Pochhammer symbol or rising factorial, 
\begin{equation}
\label{eq:ak}
(x)_k =
\begin{cases}
1 & \text{if } k=0 \\
\prod\limits_{j=0}^{k-1} x+j & \text{if } k > 0
\end{cases}.
\end{equation}

\begin{figure*}
\plotone{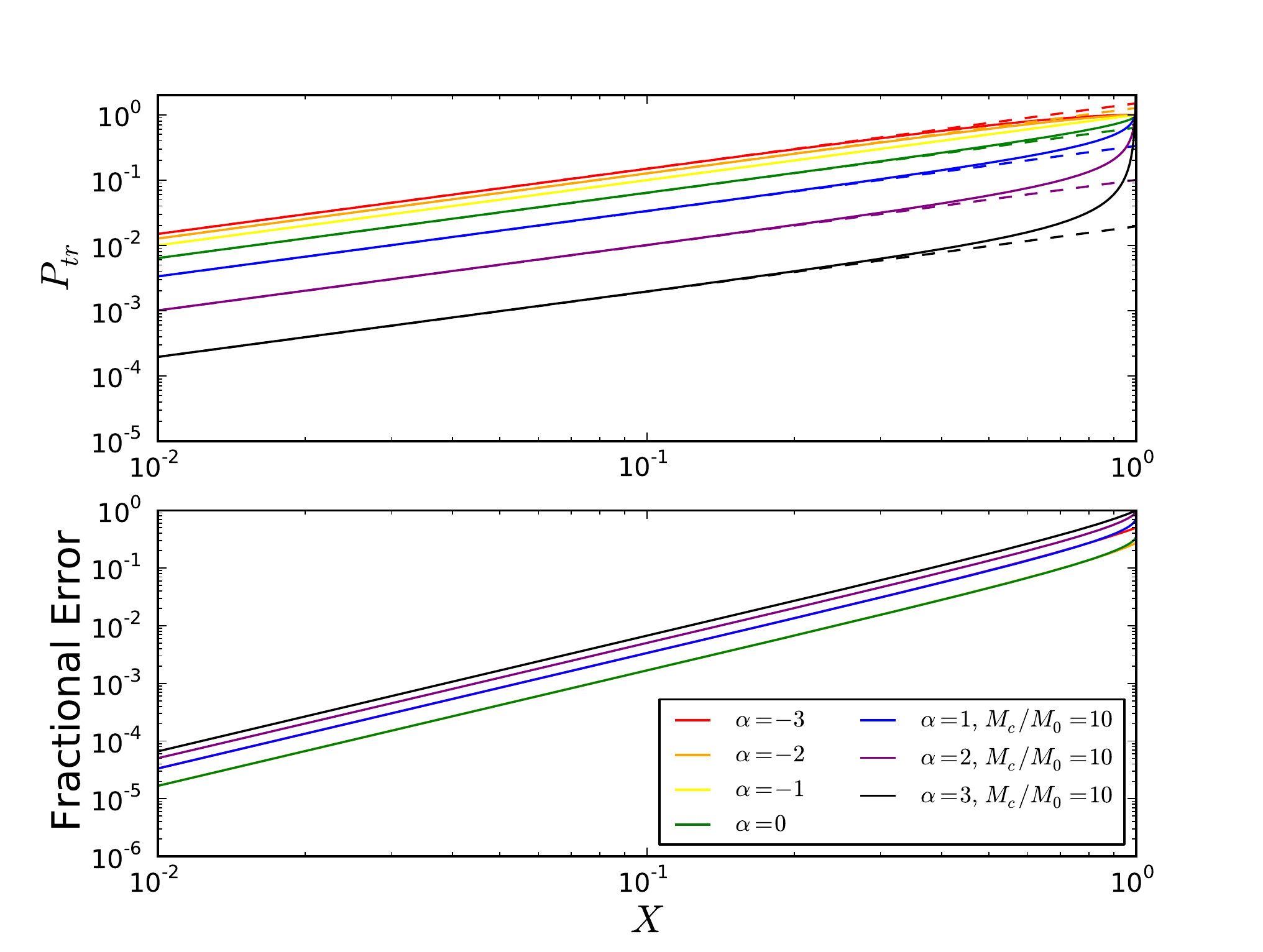}
\caption{\label{fig:probplots1} Posterior transit probability
distributions as a function of the prior transit probability $X$ for power-law true mass
distributions with indices $\alpha = -3,-2,-1,0,1,2,\ \rm{and}\
3$. For $\alpha \geq 1$, we set $\mcmax/M_0=10$ or $\sin i_{min} = 0.1$ since the
denominator of Equation \eqref{eq:tprob} does not converge for these values of 
$\alpha$.   For $\alpha = -1$, the posterior is the same as the prior distribution, so
$P_{tr} = X$.  The solid lines show the exact posterior, while the dashed
lines show the first-order Taylor series expansion centered at $X=0$. 
For small $X$, the posterior transit probability is approximately
a constant multiple of the prior probability.  The bottom panel shows the fractional deviation between
the full and approximate probabilities.}
\end{figure*}

The transit probability in equation \ref{eq:tprob} can be written
using analytic functions for specific values of $\alpha$.  Table
\ref{tab:funcform} gives these functions for integer values of
$\alpha$ between -3 and 3, and the solid lines in the top panel of
Figure \ref{fig:probplots1} shows $P_{tr}$ versus $X$ for these
$\alpha$ values.  Our posterior distributions for $\cos i$ agree with the $M_c$ posteriors given $M_0$ of \citet{Ho&Turner2011} save for $\alpha = 1$; using their notation for the numerator of the companion-mass posterior, $\Phi(M_0,X,\alpha)$, we find that $\Phi(M_0,X,1) = \log(\sqrt{X^2-1} + X)$ instead of $\Phi(M_0,X,1) = \log(2\sqrt{X^2-1} + X)$ (cf. Equation 19 of \citealt{Ho&Turner2011}).

Note that when $x$ is a negative integer, $(x)_k$ is
zero if $k\ge -x+1$.  The hypergeometric
function in Eq.\ \ref{eq:2F1} thus has $-b+1$ terms when $b$ is a negative
integer.   Therefore, when $\alpha$ is an
odd negative integer (i.e. $\alpha=-1, -3, -5, ...$), the transit
probability can be written as an odd polynomial in $X$ of order $|\alpha|$ with
$(-\alpha+1)/2$ terms.  For $\alpha=-1$, the hypergeometric functions
in the numerator and denominator of Equation \ref{eq:hyperprob} are
unity, and thus the transit probability is simply $X$ --- i.e. the
prior transit probability.

Smaller $\alpha$ corresponds to a prior increasingly weighted toward
smaller values of $M_c$.  For a given $M_{0}$, $M_{c} = M_{0}/\sin i
\propto 1/\sin i$, so larger $\sin i$ values are increasingly preferred for
smaller $\alpha$. As a result, smaller $\cos i$ values are more
likely, and the transit probability is higher, as illustrated in
Figure \ref{fig:probplots1}.

\begin{figure*}
\plotone{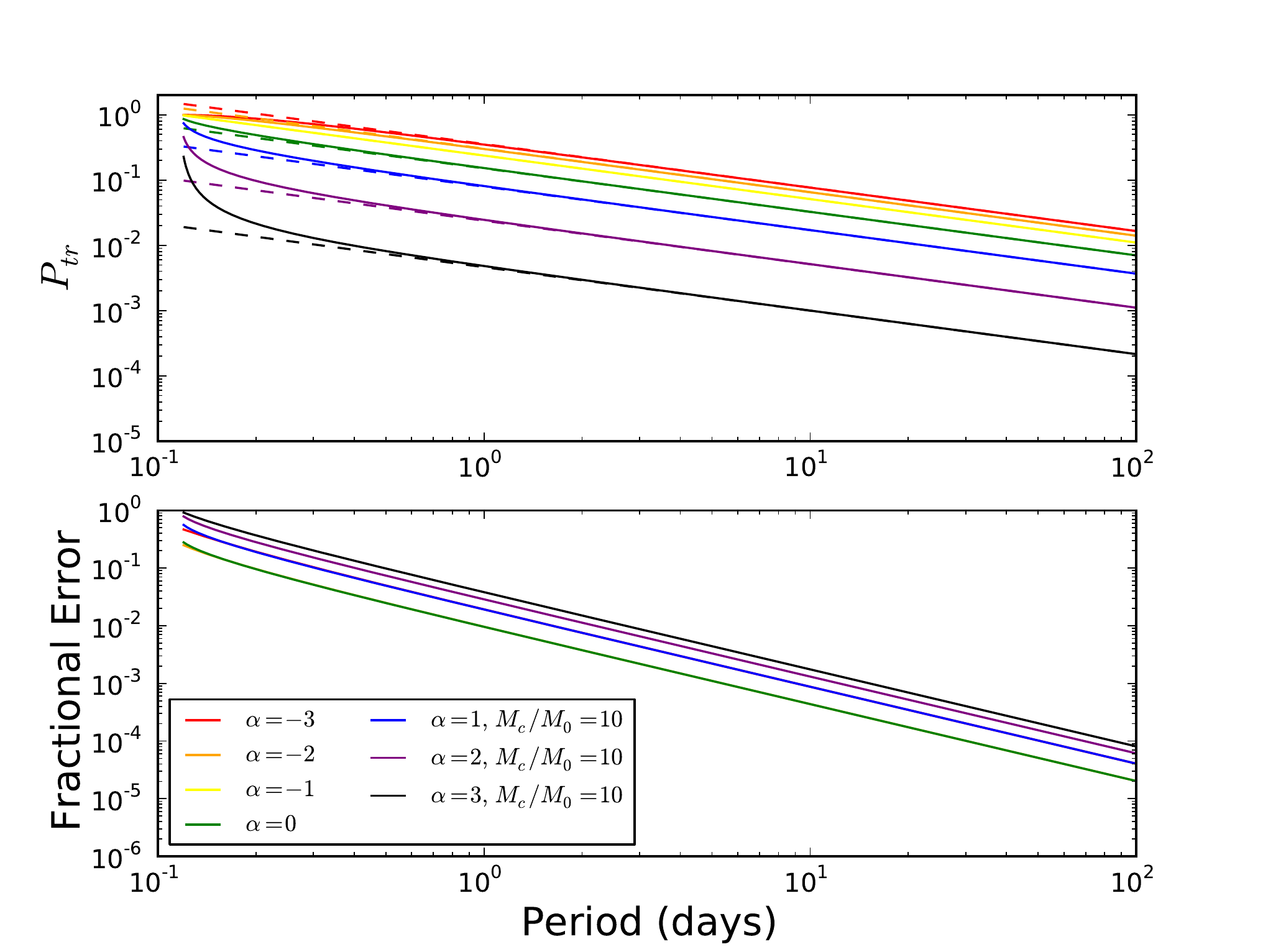}
\caption{\label{fig:probplots2} The top panel shows the posterior
transit probability distributions as a function of period (solid
lines) and their corresponding first-order Taylor approximations
(dashed lines) for integer values of $\alpha$ between -3 and 3. We
assume a solar mass and radius. For a given period, the top panel
indicates that the posterior transit probability can differ by nearly
a factor of $10^2$ for different $\alpha$. The difference in transit
probabilities for exceptionally short periods decreases rapidly with
decreasing period, converging to 1 for $T \rightarrow 0$. The bottom
panel shows the fractional deviation of the first-order Taylor
approximations to the full analytical transit probabilities.  
}
\end{figure*}

We can also express Equation \eqref{eq:tprob} as a function of period
with Equation \eqref{eqn:semi}. We assume that $M_*=M_\odot$, $R_*=R_\odot$,
and $R_c \ll R_*$.  
The top panel of Figure \ref{fig:probplots2} shows
the posterior transit probabilities for the same power-law true mass
distributions as a function of period. The dependence on the power-law
true mass distribution can create a factor of $\sim 10^2$
difference in the posterior transit probability at a given period for 
power-law exponents in the range $\alpha=-3$ to $3$.  
This disparity decreases for very small orbital periods since
$P_{tr} \rightarrow 1$ as $T \rightarrow 0$.

\subsection{Dependence on Maximum $M_c$ for $\alpha \geq 1$}\label{subsubsec:Convergence}

\begin{figure*}
\plotone{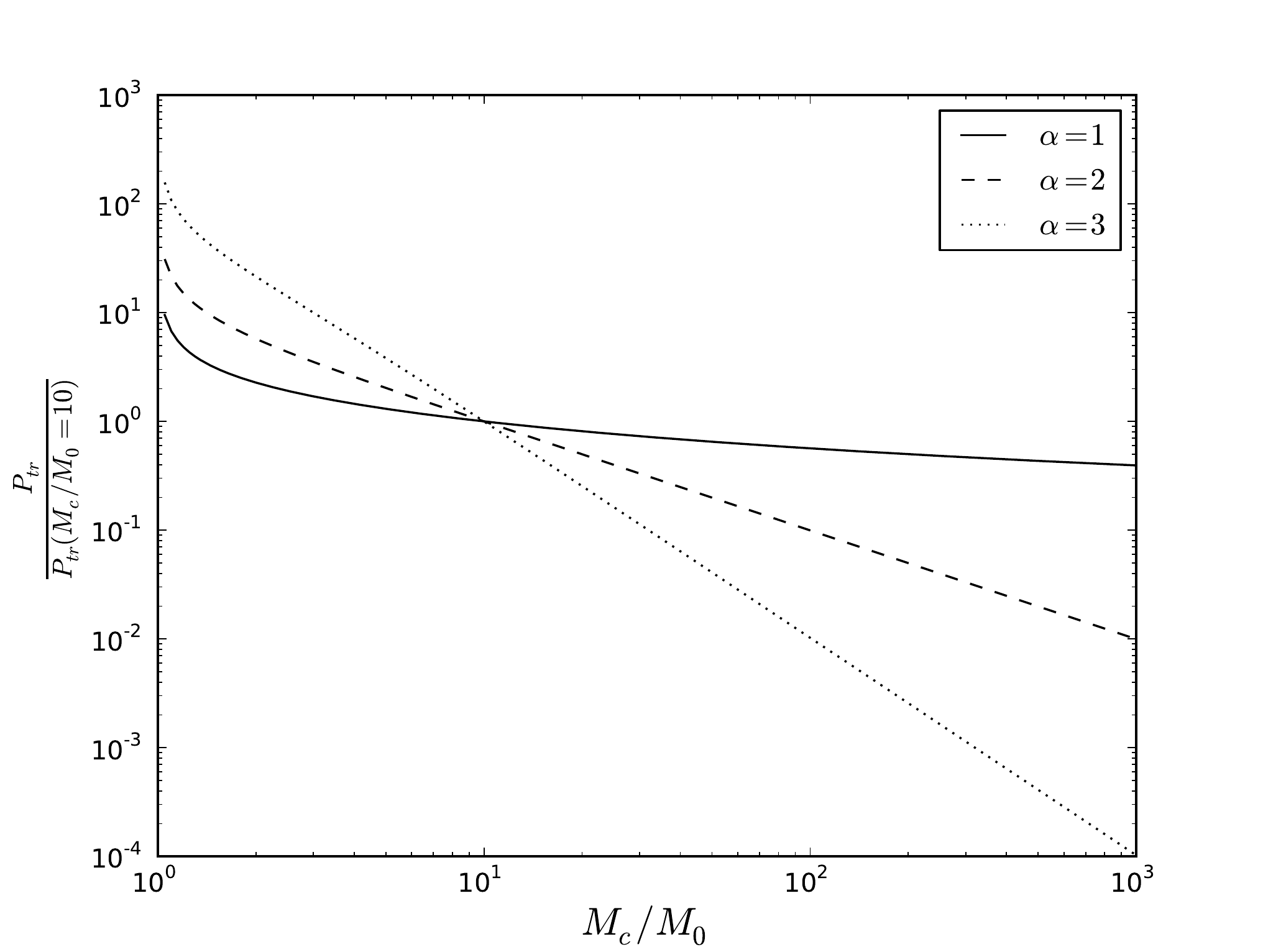}
\caption{\label{fig:convergence} Posterior transit probabilities for
$\alpha$ = 1, 2, and 3 as a function of $M_c/M_0$, normalized to
the corresponding transit probability with $M_c/M_0 =
10$. Increasing $M_c/M_0$ corresponds to increasing $\cos
i_{min}$. $P_{tr}$ diverges only logarithmically for $\cos i_{min}$
approaching unity for $\alpha=1$, but diverges more rapidly for
$\alpha > 1$, as illustrated for the specific cases $\alpha = 2$
and 3.}
\end{figure*}

As noted previously, the denominator of Equation \eqref{eq:tprob} does
not converge for $\cos i_{min}=1$ when $\alpha \geq 1$. In Figure \ref{fig:convergence} we
show the sensitivity of the posterior transit probability to $\cos i_{min}$ for
$\alpha$ = 1, 2, and 3.  We normalized $P_{tr}$ to its value assuming $\mcmax/M_0 = 1(\sin
i_{min} = 10$.  Since $M_c/M_0 = \frac{1}{\sin i} = (1 - \cos^2
i)^{-1/2}$, increasing $\mcmax$ corresponds to
increasing $\cos i_{min}$, and therefore lowering the transit probability. 
For $\alpha=1$, $P_{tr}$ diverges logarithmically, and therefore the precise
choice of $\mcmax$ does not strongly affect the implied transit probability,
for reasonable values. 

On the other hand, $P_{tr}$ diverges more strongly for $\alpha > 1$,
and in particular depends sensitively on $\mcmax$, and can be
arbitrarily small.  Accurate estimates of the transit probability of
companions with minimum masses in such steeply-rising portions of the
companion mass function therefore require reasonably narrow
constraints on $\mcmax$.  In some cases, other evidence can be used to
place a limit on $\mcmax$.  For example, companions around very
massive main-sequence stars can often be ruled out based on the lack
of evidence of flux from the companion or a second set of spectral
lines (e.g., \citealt{fleming2012}).

\subsection{Taylor Approximations}\label{subsec:Taylor}
The simple linear behavior for small values of $X$ exhibited in
Figures \ref{fig:probplots1} and \ref{fig:probplots2} invites us to
make a first-order Taylor approximation for $P_{tr}$ centered at $X =
0$. The numerator of Equation \eqref{eq:hyperprob} can be written as
\begin{eqnarray}
\label{eq:numgamma}
X\left[_2F_1\left(\frac{1}{2},\frac{\alpha + 1}{2};\frac{3}{2};X^2\right) \right] &=& X\left[\sum\limits_{k=0}^\infty\frac{\left(\frac{-[\alpha +1]}{2}\right)_k}{1+2k}\frac{X^{2k}}{k!}\right]\nonumber\\ 
&=& X[1 + \mathcal{O}(X^2)].
\end{eqnarray}
This allows us to express the Taylor approximation as
\begin{equation}
\label{eq:tprobtaylor}
P_{tr} = f_\alpha X + \mathcal{O}(X^3).
\end{equation}
In other words, the posterior transit probability for power-law priors
on the companion mass is equal to the prior transit probability times
a constant scale factor $f_\alpha$, up to third order in $X$.  The
scale factor $f_\alpha$ depends only on $\alpha$ and $i_{\min}$, and
is simply equal to the inverse of the denominator in the full
expression for the transit probability,
\begin{equation}
\label{eq:falpha}
f_\alpha \equiv \left[_2F_1\left(\frac{1}{2},\frac{\alpha + 1}{2};\frac{3}{2};\cos^2 i_{min}\right)\right]^{-1}.
\end{equation}

For $\alpha < 1$, we can use Gauss' Theorem to express $f_\alpha$ in terms of Gamma functions, 
\begin{equation}
\label{eq:scalegamma}
f_\alpha = \frac{2\Gamma(1-\frac{\alpha}{2})}{\sqrt{\pi}\ \Gamma(\frac{1-\alpha}{2})}.
\end{equation}
It can be shown that this expression diverges for $\alpha \geq
1$. Furthermore, for $\alpha \ll -1$, we can use Stirling's
approximation that $n! \approx \sqrt{2\pi n}n^ne^{-n}$ and the
identity $n!=\Gamma(n+1)$ to approximate equation
\eqref{eq:scalegamma} as
\begin{equation}
\label{eq:stirling}
f_\alpha \simeq \sqrt{\frac{-2\alpha}{\pi e}}\left(1+\frac{1}{\alpha}\right)^{\alpha/2}.
\end{equation}
An alternate expression for $f_\alpha$ in this regime that is somewhat
less accurate, but also somewhat simpler is, 
\begin{equation}
\label{eq:fapprox}
f_\alpha \sim \sqrt{\frac{-2\alpha}{\pi}}.
\end{equation}
This can be derived from either Equation \ref{eq:stirling} or
directly from the definition of $f_\alpha$ in equation
\ref{eq:falpha}, by using the definition of the exponential function, $e^x
\equiv \lim_{n\to \infty} (1+x/n)^n$.  

\begin{figure*}
\plotone{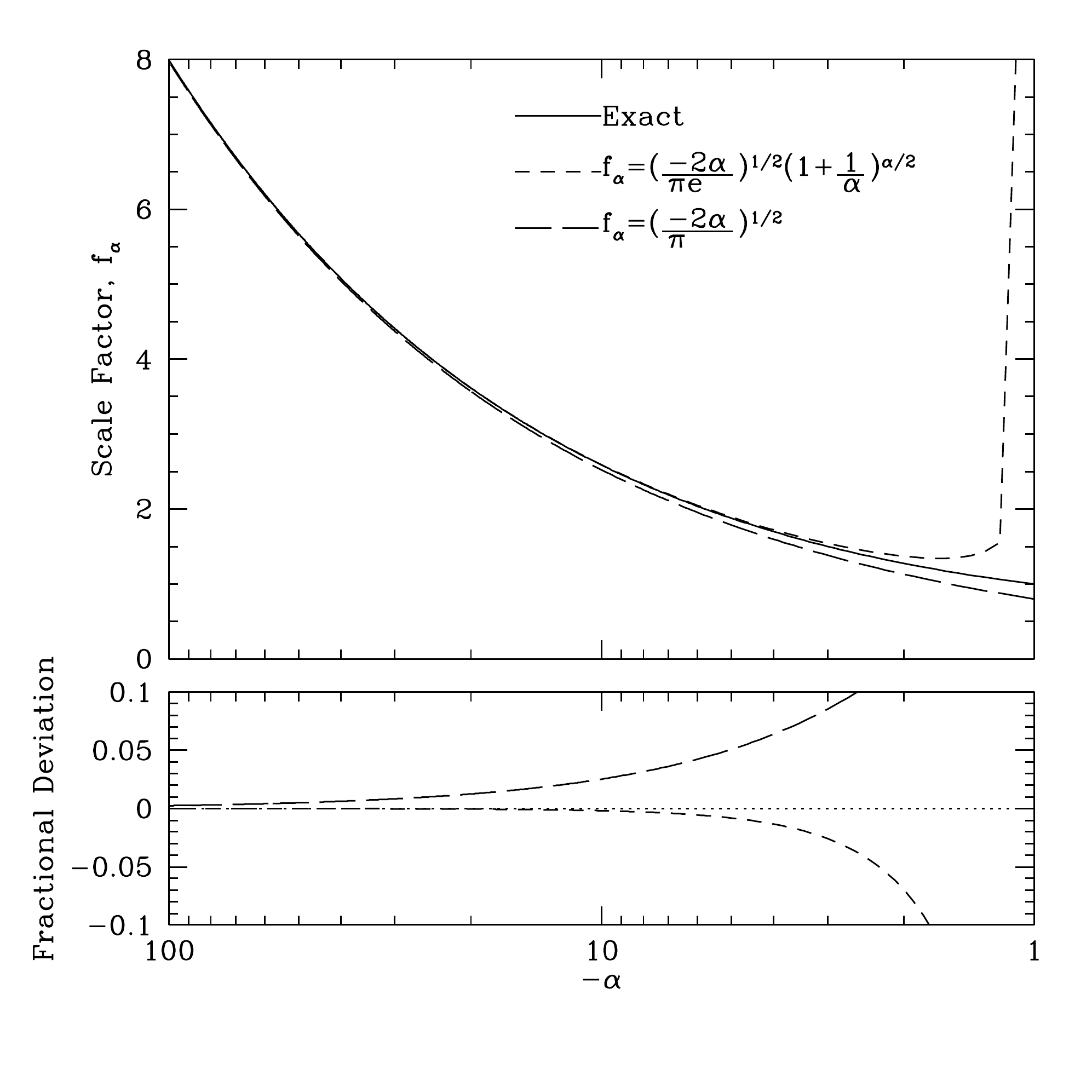}
\caption{\label{fig:falpha} \emph{Top panel}: The first-order posterior transit scaling
factor $f_\alpha$ as a function of $\alpha$ for negative values of
$\alpha$.  The solid line shows the exact expression obtained from
numerical integration of Equation \ref{eq:falpha} with $\cos i_{min}=1$.
The short-dashed line shows the approximation to $f_\alpha$
in Equation\ \ref{eq:stirling}, whereas the long-dashed line
shows the simpler approximation in Equation \ref{eq:fapprox}.
\emph{Bottom panel}: The fractional deviation of the approximations to $f_\alpha$
from the exact expression as a function of $\alpha$.}
\end{figure*}
\begin{table*}
\begin{center}
\caption{\label{tab:funcform} Posterior Transit Probabilities and Scale Factors}
\scriptsize
\begin{tabular}{cccccc}
\tableline
\multirow{2}{*}{Power-law Index $\alpha$} & \multirow{2}{*}{$P_{tr}(X)$} & \multirow{2}{*}{$f_\alpha$} & \multicolumn{3}{c}{Max $X$ for \% Error} \\\cline{4-6}
& & & $\leq 1\%$ & $\leq 5\%$ & $\leq 10\%$ \\
\tableline
-3 & $\frac{3}{2}(X - \frac{X^3}{3})$ & 3/2 & 0.17 & 0.38 & 0.52\\
-2 & $\frac{2}{\pi}(X\sqrt{1-X^2}+ \arcsin X)$ & $4/\pi$ & 0.24 & 0.52 & 0.71\\
-1 & X & 1 & -- & -- & --\\
0 & $\frac{2}{\pi} \arcsin X$ &$2/\pi$ & 0.24 & 0.52 & 0.71 \\
1 & $\frac{\arctanh X}{\arctanh( \cos i_{min})}$ & $[\arctanh (\cos i_{min})]^{-1}$ & 0.17 & 0.38 & 0.53\\
2 & $\frac{X}{\sqrt{1 - X^2}}\tan i_{min}$ & $\tan i_{min}$ & $0.14$ & $0.31$ &$0.44$\\
3 & $\frac{\arctanh X - \frac{X}{X^2 - 1}}{\arctanh(\cos i_{min}) + \cot i_{min} \csc i_{min}}$ & $2[\arctanh (\cos i_{min}) + \cot i_{min} \csc i_{min}]^{-1}$ & 0.12 & 0.27 & 0.38\\
\tableline
\end{tabular}
\end{center}
\end{table*}
Figure \ref{fig:falpha} shows the exact value of $f_\alpha$ for $\alpha \le 0$, as
well as the two approximations given in Equations \ref{eq:stirling}
and \ref{eq:fapprox}.  We find that Equation \ref{eq:stirling} is an
excellent approximation to $f_\alpha$, deviating by $<10 \%$ for $\alpha
\le -1.7$ and by $<1\%$ for $\alpha \le -4.4$.  Equation
\ref{eq:fapprox} is a somewhat poorer, but still accurate,
approximation, deviating by $\la 10\%$ for $\alpha \le -2.5$.
Interestingly, the boost factor does not converge for $\alpha
\rightarrow -\infty$, and thus the posterior transit probability can
be arbitrarily large even for small prior transit probabilities, if one
allows for a mass function that rises arbitrarily steeply toward small
masses.  This implies that detected RV companions with minimum masses
near a large jump in the mass function toward lower masses will have
quite high transit probabilities.

The exact scale factors for several values of $\alpha$ are given in Table
\ref{tab:funcform}. We find the transit probability is boosted by $\sim 30\%$ for 
$\alpha=-2$,  and by $50\%$ for $\alpha=-3$.\footnote{We note that $f_{-5}=15/8 \sim 1.9$, 
$f_{-7}=35/16\sim 2.2$, $f_{-9}=315/128 \sim 2.5$.} However, the transit probability is reduced
by $\sim 64\%$ for $\alpha=0$ ($f_0= 2/\pi$), and by larger amounts for $\alpha > 0$.

The linear approximations to $P_{tr}$ are plotted as dashed lines in the top panels of
Figures \ref{fig:probplots1} and \ref{fig:probplots2}, and the bottom
panels show the fractional error between the full analytical
expressions and our Taylor approximations. 
The bottom panels indicate
that the approximations are quite accurate for all but the smallest period orbits.
In order to provide a more quantitative estimate of how well the Taylor expansions
approximate the true posterior expressions, we provide in 
the right columns of Table \ref{tab:funcform}
the values of prior transit probability $X$ (or $R_*/a$ for circular orbits) for which the
Taylor approximations deviate from the analytical expressions for
$P_{tr}$ by 1\%, 5\%, and 10\%. For $\alpha \leq 0$, the approximate
transit probability expressions are accurate to within 10\% for most
values of $X$; the 10\% accuracy threshold moves to lower $X$ with
increasing $\alpha$. For the eight transiting planets originally discovered
by RV that were known at the
time of publication, $X$ ranges from $\sim 0.015$ for GJ 436 b to
$\sim 0.29$ for 55 Cnc e, so the approximate transit probabilities for
these planets have $<10\%$ error (in fact, the approximations for all
but 55 Cnc e are within a $5\%$ error). If we can tolerate errors of one
part in 10, we can confidently use these approximate posterior
expressions for most situations. 

\section{Posterior Transit Probabilities for Model Planet Mass Distributions}\label{sec:bd}
While the results from the previous section hold for power-law distributions,
the true mass distribution of companions to stars of any type is not
so well-behaved. For example, the brown dwarf desert
is a well-known feature in the mass function of relatively
short-period companions to solar-type stars
\citep{marcy2000}.  It represents a local minimum in this
mass function, with the frequency of companions rising toward the
stellar regime on the high-mass side of the desert and toward the
planetary regime on the low-mass side of the desert
\citep{GretherLineweaver2006}.

The biggest obstacle to providing robust a posteriori transit
probabilities is that the mass distribution of planetary companions is
poorly known.  Radial velocity detections yield only the minimum mass
of the companion, whereas planets discovered from ground-based transit
surveys have severe and difficult-to-quantify selection biases \citep{pont2005,gaudi2005a,gaudi2005b,fressin2007}.  While
Kepler has dramatically increased the number of known transiting
planets, we do not have mass measurements for most of these systems.

We therefore adopt synthetic, but plausible, physically-motivated
planetary companion mass distributions that are derived from detailed
planet formation simulations.  We emphasize that we use these simply
to illustrate the effects of more complicated mass functions on the
transit probability; we do not claim to provide firm or robust
predictions for these probabilities based on these mass functions.

We consider mass functions predicted by the planet formation models of
\citet{IdaLin2008} and \citet{Mordasini2012}.  To each of these, we
also add stellar companions using a simple model based on the
compilation of multiple stellar systems of \citet{Raghavan2010}.  Since the
aridity of the brown dwarf desert is poorly constrained, we also
consider two extreme variants of each mass function: one with brown dwarf
companions with mass between the deuterium-burning and the
hydrogen-burning limit as predicted by the planet formation theories, 
and one without such companions.  We assume a
deuterium-burning limit of $M \gtrsim 13M_{\rm Jup}$ \citep{spiegel2011} and a
hydrogen-burning limit of $M \gtrsim 0.07 M_\odot$ \citep{chabrier2000}.  We also assume
all companions are on circular orbits.  We discuss the determination
of these mass distributions and analysis of the resulting posterior
transit probability distributions in detail below.

\subsection{Discussion of the True Mass Distributions}\label{subsec:bddist}
We took true masses of sub-stellar companions from the population
synthesis models of \citet{IdaLin2004} and
\citet{Mordasini2012}. Although the companion mass function in these
models depends on the period, for simplicity we included all planetary
companions, regardless of period.

For stellar-mass companions, we adopted a distribution based on the
results of \citet{GretherLineweaver2006} and
\citet{Raghavan2010}. Both groups examined a volume-limited sample of
companions around Sun-like stars out to 25 pc from the
\emph{Hipparcos} catalog. Out to 25 pc, the mass ratio of the
companions around Sun-like stars, $M_c/M_{*}$, is nearly flat over the
linear mass ratio range 0.2-0.95 \citep{Raghavan2010}. Assuming that the host
stars are all of solar mass, then the distribution of companion masses
is approximately flat in $M_c$ over the range
$0.2\textrm{M}_\sun-0.95\textrm{M}_\sun$. For simplicity, we set our
mass distribution to be flat in $M_c$ between $0.08\textrm{M}_\sun$ and $0.95\textrm{M}_\sun$.

We then combined the planetary and stellar mass distributions by
normalizing the different mass regions, assuming that the
planetary-mass companions from the model distributions with
$1\textrm{M}_{\rm Jup} \lesssim M_c \leq 13 \textrm{M}_{\rm Jup}$ are twice as
common as stellar-mass companions around Sun-like stars
\citep{GretherLineweaver2006}.  These combined mass distributions
contain companions predicted by the planet formation models with
masses in the brown dwarf regime.  We also consider mass distributions
with a completely arid brown dwarf desert.  For these, we remove
companions with masses between $13\textrm{M}_{\rm Jup}$ and
$0.07\textrm{M}_\sun$. Thus, we consider four different mass distributions:
two with masses from \citet{IdaLin2004} and two with masses from
\citet{Mordasini2012}, with one of each having a completely dry brown
dwarf desert.

\begin{figure*}
\plottwo{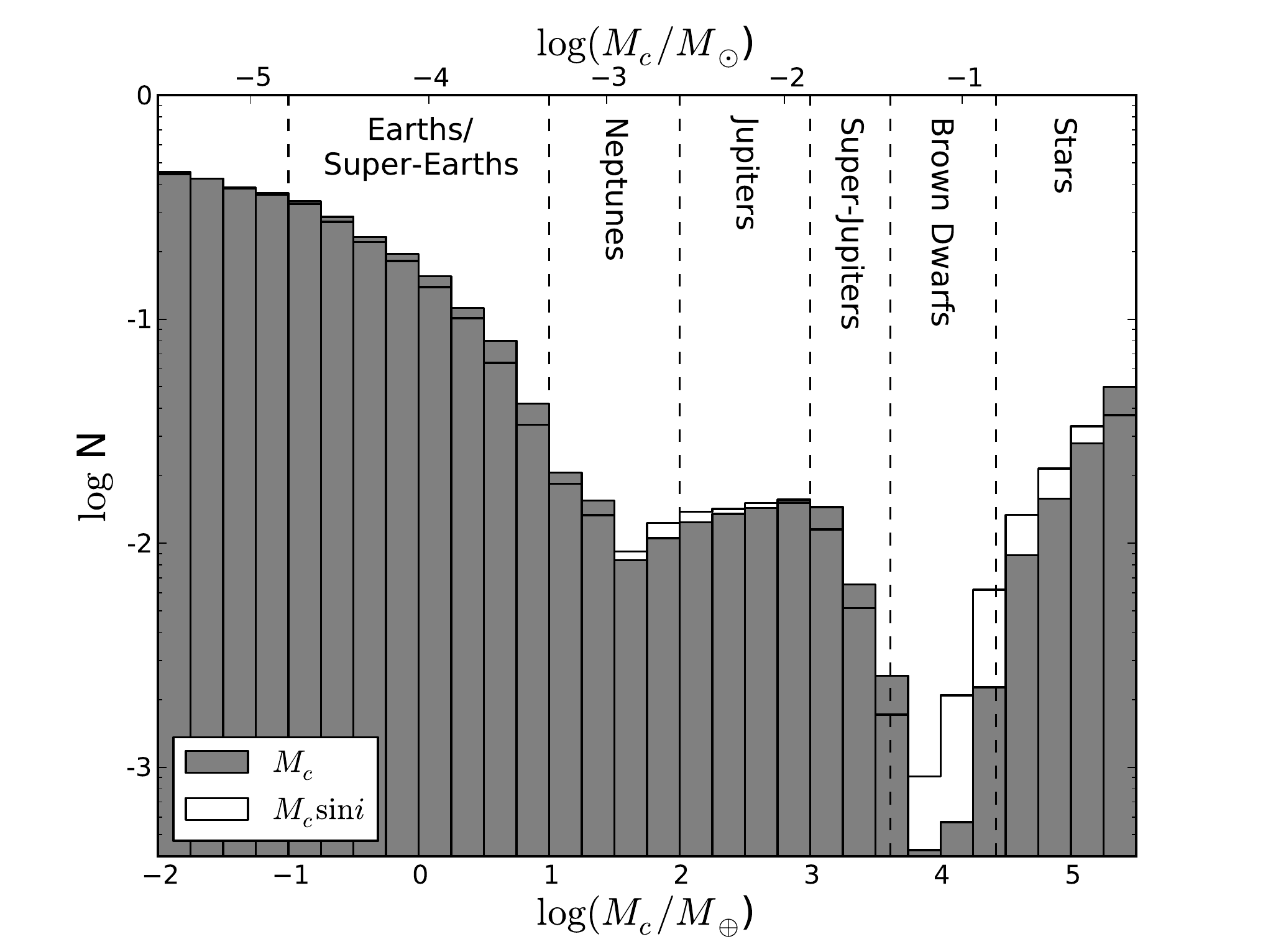}{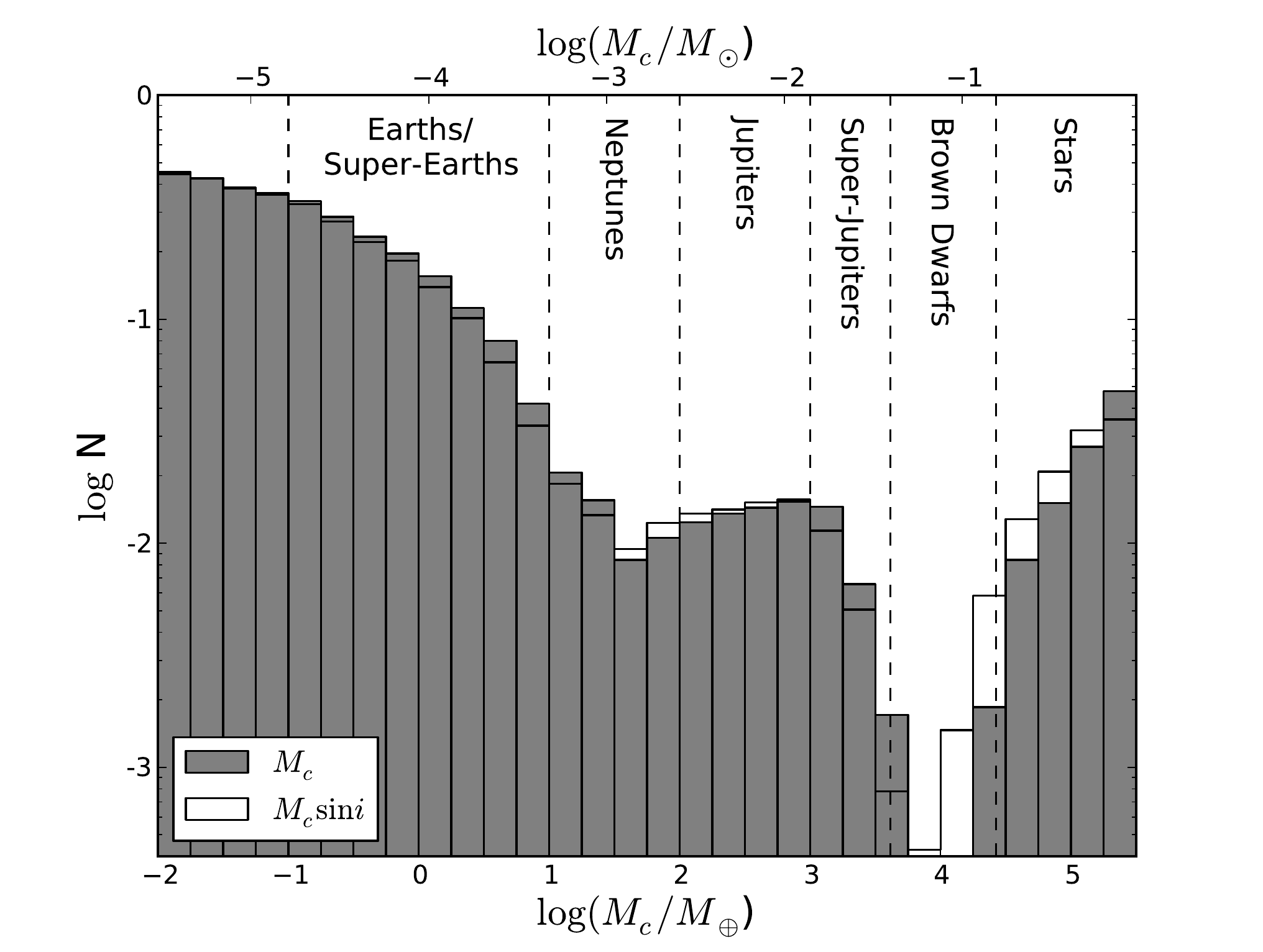}
\plottwo{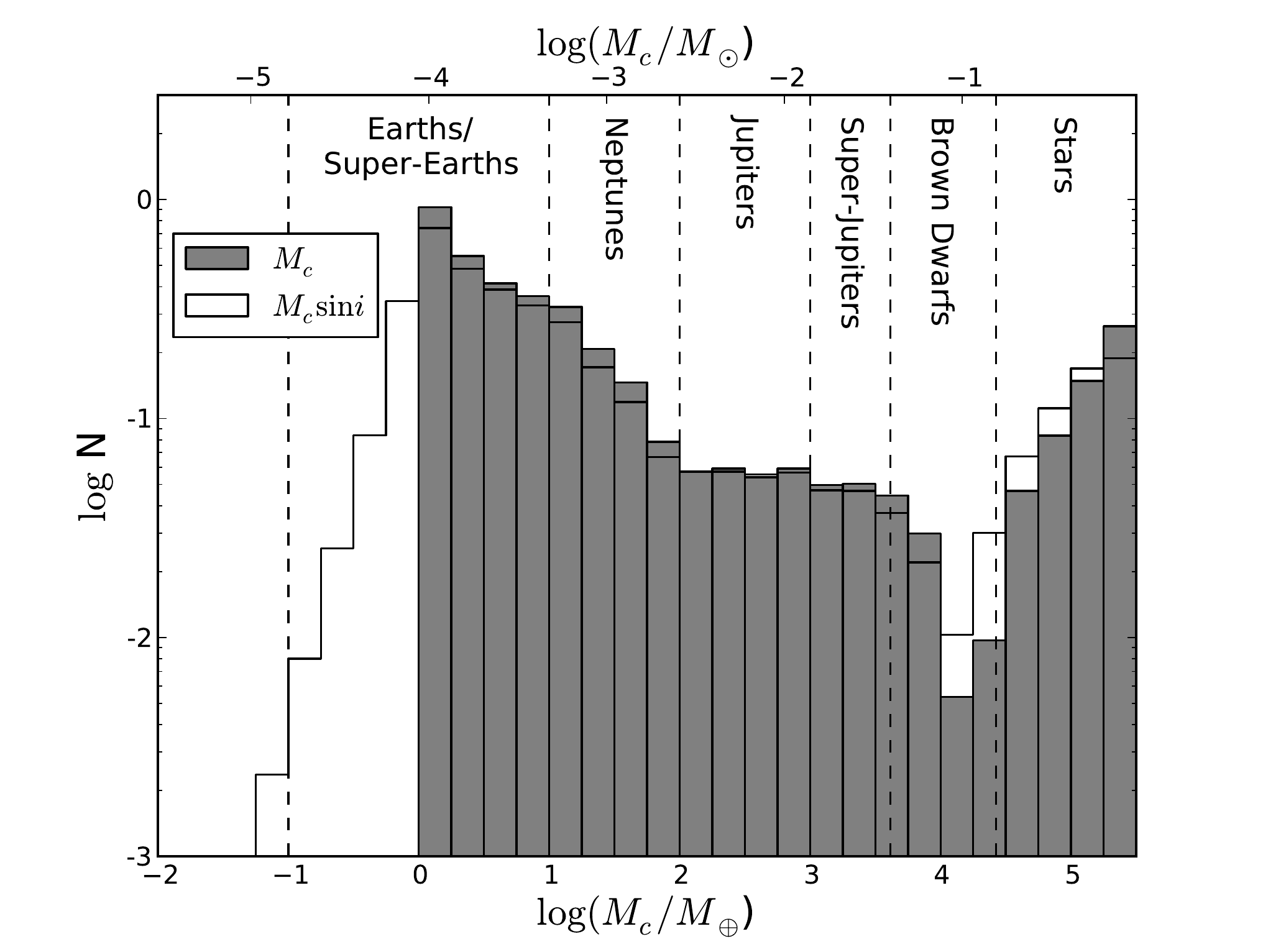}{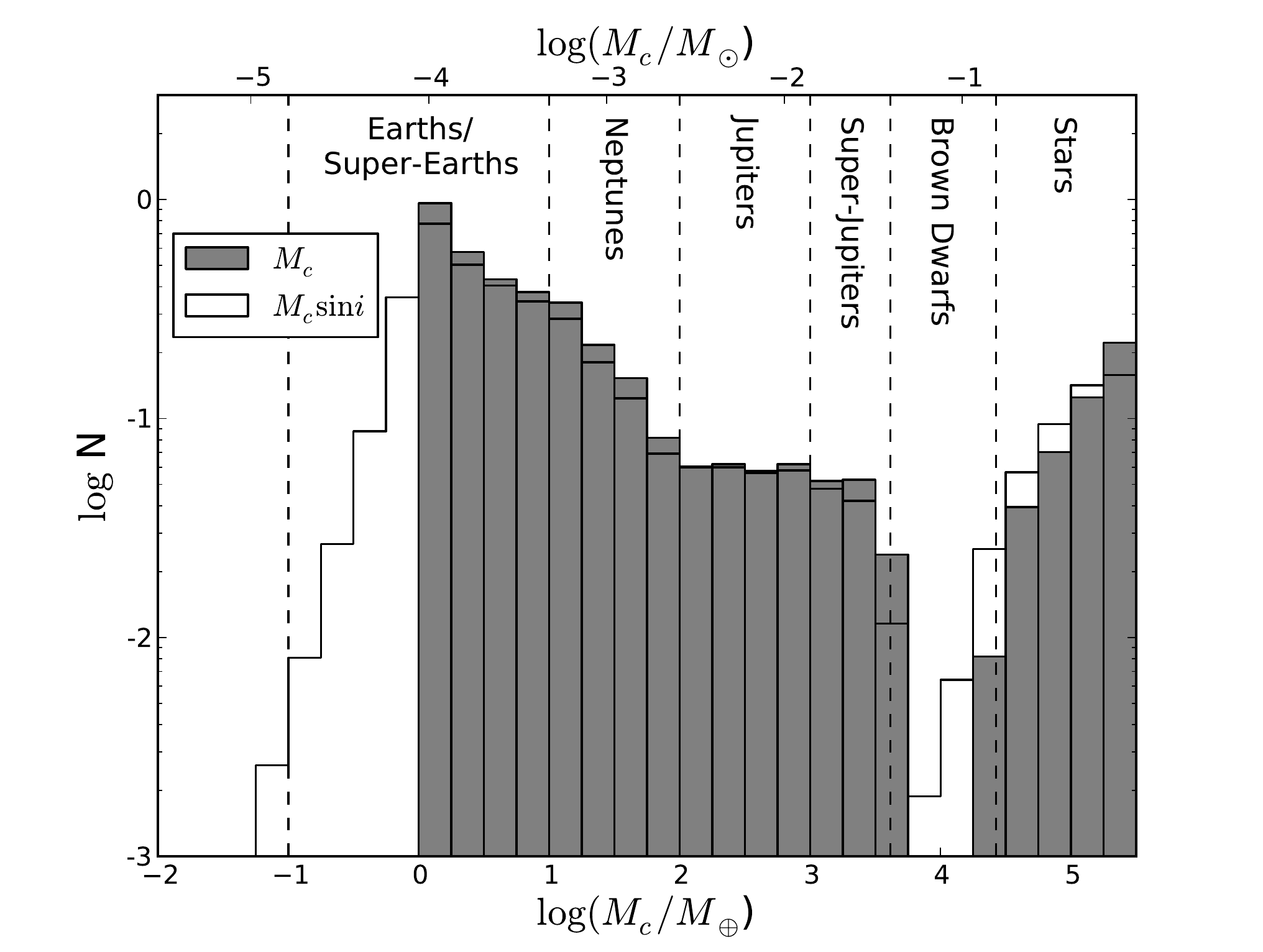}
\caption{\label{fig:massdist} Mass distributions of companions around
Sun-like stars using planetary companions from Ida \& Lin (top) and
Mordasini (bottom), with (left) and without (right) brown
dwarf companions in the deuterium-burning to hydrogen-burning mass
range. The true mass distributions are shown as filled histograms,
whereas the minimum mass distributions are outlined. The
vertical dashed lines divide the plots into the sub-Earth,
Earth/Super-Earth, Neptune, Jupiter, Super-Jupiter, Brown Dwarf and
stellar mass regimes as described in the text. Values of $\alpha$ for
the various regimes are as follows, starting with Earths/Super-Earths
and increasing in mass: $\alpha \approx -1.5, -1.4, -0.9, -2.2, -0.9,
\textrm{and} -0.2$ for the top-left panel; $\alpha \approx -1.5, -1.4,
-0.9, -2.5, -0.8, \textrm{and} -0.1$ for the top-right panel; $\alpha
\approx -1.4, -1.7, -1.1, -1.1, -1.5,$ and $-0.1$ for the
bottom-left panel; and $\alpha \approx -1.4, -1.7, -1.1, -1.6, -1.3,$ and $-0.1$ for the bottom-right panel. }
\end{figure*}

For each of these $M_c$ distributions, we created a distribution of
minimum masses $M_0$ as follows: For each true mass, we drew several
values of $\cos i$ from a uniform distribution, transformed these to
$\sin i$, and then multiplied each true mass by the $\sin i$ values to
obtain minimum masses.  Figure \ref{fig:massdist} show the final
distributions of true and minimum masses, binned in in 0.25 dex bins in log mass.
We show the distributions obtained from the two planet formation models, each
with and without a completely dry brown dwarf desert.
To guide the discussion, we have subdivided the mass
function into several (somewhat arbitrary) regimes: sub-Earths
($10^{-8}\textrm{M}_\earth-0.1\textrm{M}_\earth$), Earths/Super-Earths
($0.1\textrm{M}_\earth-10\textrm{M}_\earth$), Neptunes
($10\textrm{M}_\earth-100\textrm{M}_\earth$), Jupiters
($100\textrm{M}_\earth-10^{3}\textrm{M}_\earth$), Super-Jupiters
($10^{3}\textrm{M}_\earth-13\textrm{M}_{\rm Jup}$), brown dwarfs
($13\textrm{M}_{\rm Jup}-0.07\textrm{M}_\sun$) and stellar companions
($0.07\textrm{M}_\sun-1\textrm{M}_\sun$).

The distributions all appear qualitatively similar and have several
common features worth noting.  In the stellar regime, the mass
function falls toward lower masses as a power-law with a slope
$\alpha \sim 0$ (by design).  There is a local minimum in the brown dwarf
regime, with the mass function rising toward lower masses continuing
into the planetary regime.  In particular, there is a relatively sharp
rise for Super-Jupiters, with the mass function roughly behaving as a
power law with $\alpha \sim -2$ to $\alpha \sim -2.5$ in this regime.  The mass function for
Jupiters is essentially flat in $\log(M_c)$, i.e., $\alpha \sim -1$.
For lower-mass planets, the mass function again begins to rise, with
$\alpha \sim -1.5$ from the Neptune through the
Earth/Super-Earth regime.  Finally, the planet formation models
predict a peak in the mass function at or below an Earth mass, with a
fall off for companions less massive than this peak.

\subsection{Posterior Transit Probability Calculations and Results}\label{subsec:bdprob}

We can now use the companion $M_0$ values in the aforementioned mass
distributions to estimate the posterior transit probabilities for
planets in bins of $M_0$.  To proceed, we assume that the host star is
a solar radius and that the companion orbits with a semimajor axis of
0.1 AU.  We then calculate the transit probability of each $M_0$ bin
by examining the fraction of companions in that bin that transit ---
i.e.  the fraction for which $\cos i \leq R_{\sun}$/(0.1 AU).  We then
normalized the transit probability to the a priori transit probability
$R_{\sun}$/(0.1 AU), thereby providing an estimate of the factor $f$
by which the a posteriori transit probability is scaled relative to the prior transit
probability, in analogy to $f_\alpha$ for the power-law mass function
priors.
\begin{figure*}
\plottwo{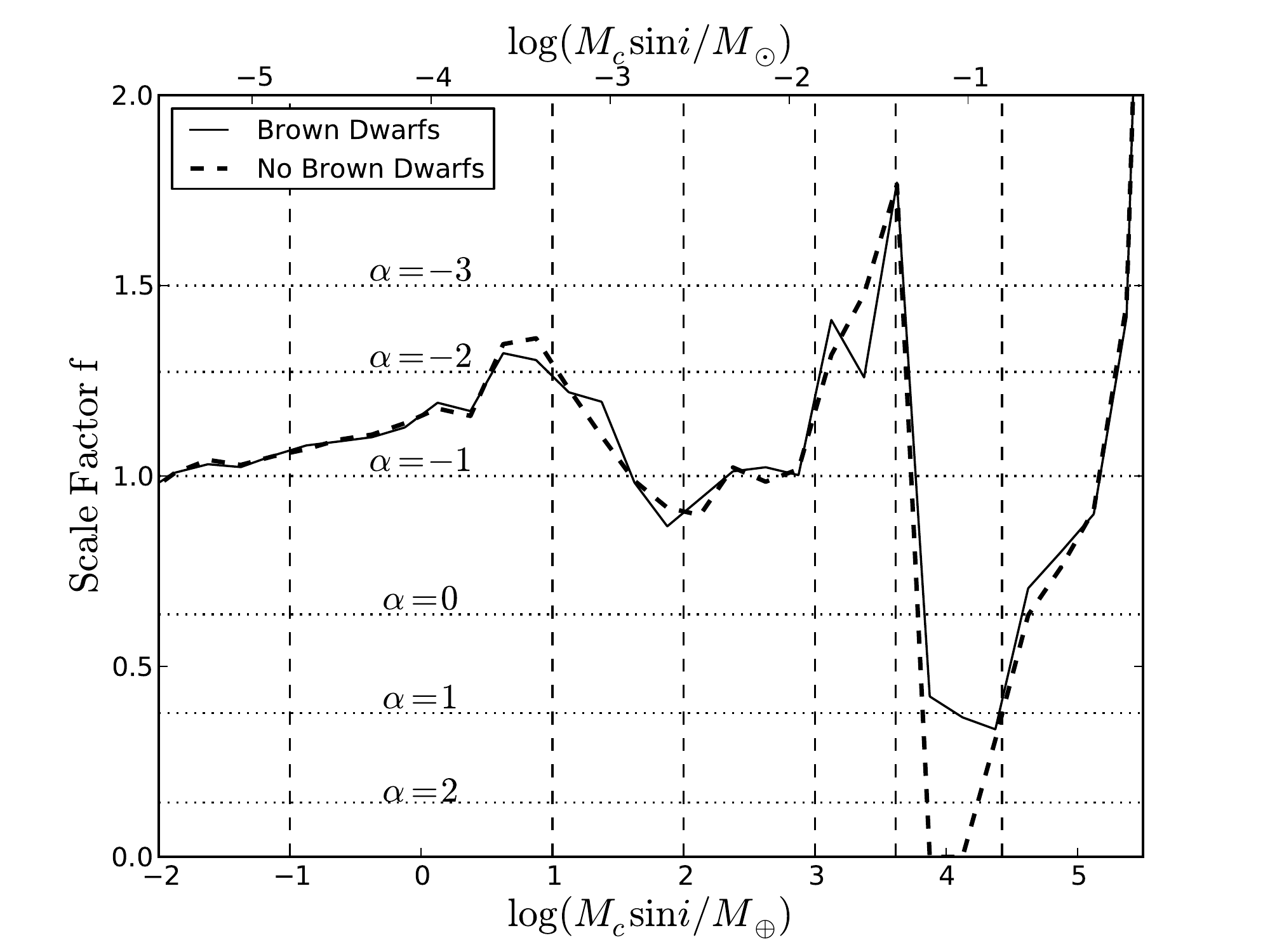}{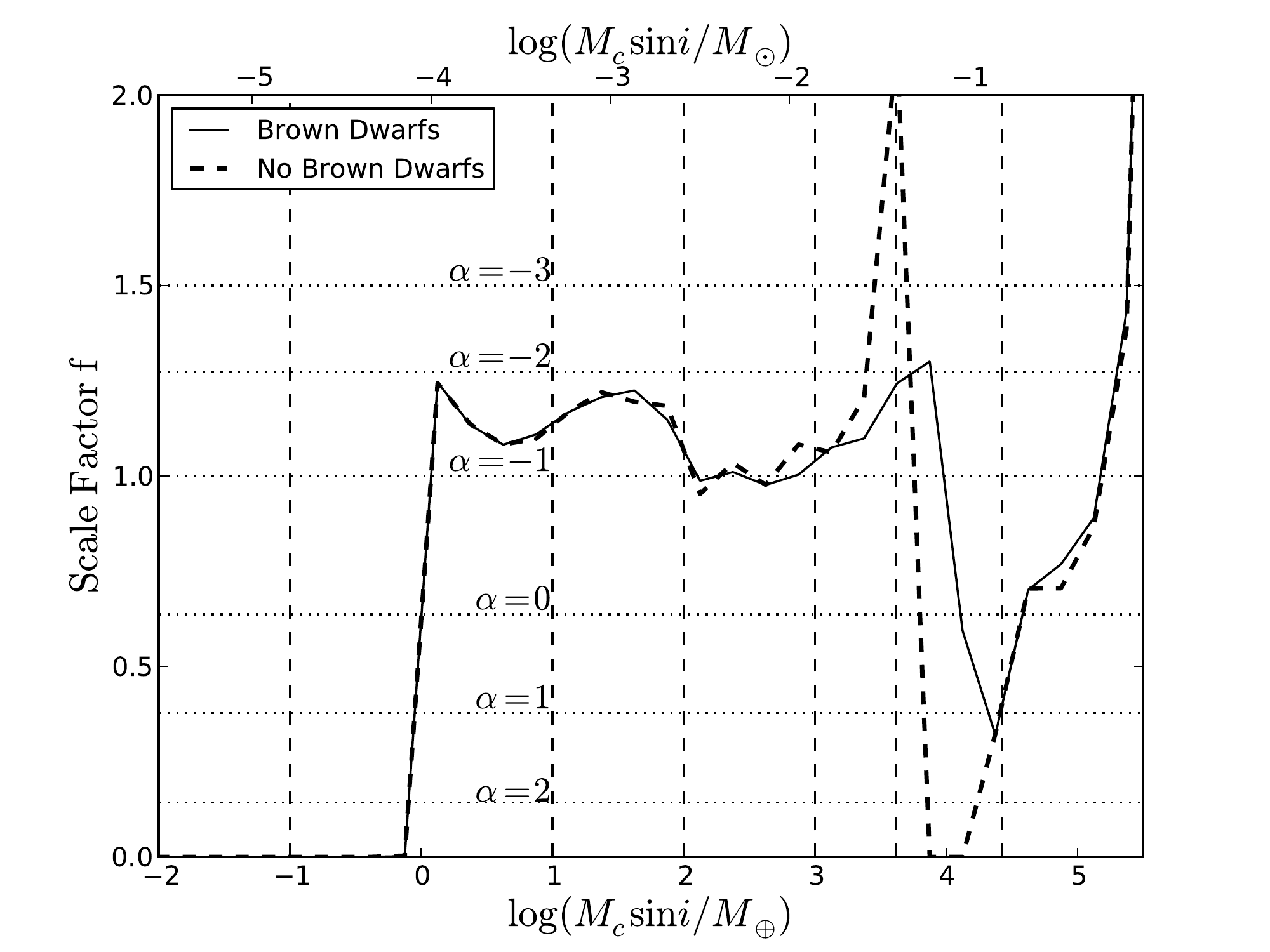}
\plottwo{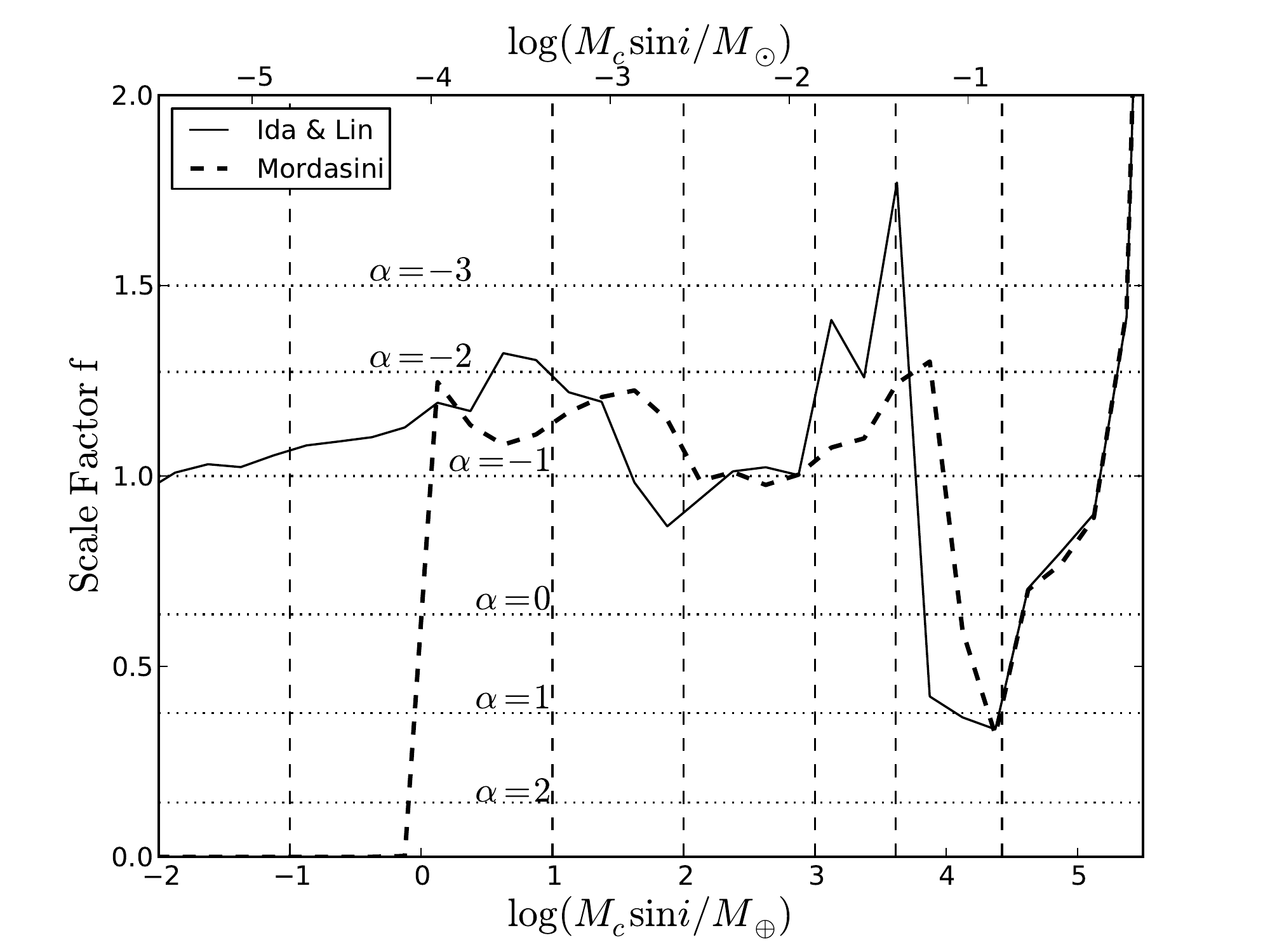}{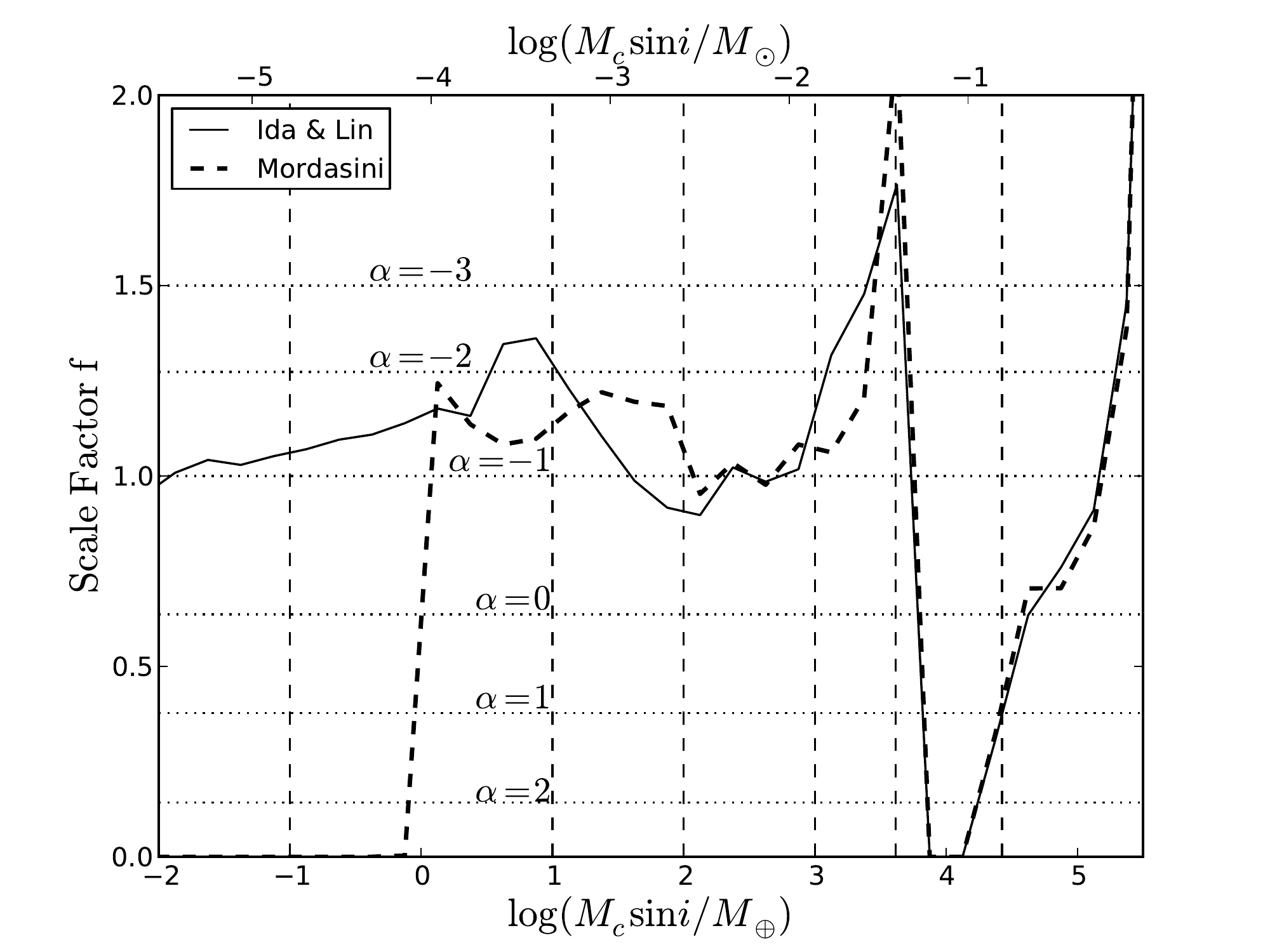}
\caption{\label{fig:Ptformasses}Scale factors $f$ (ratio of posterior
to prior transit probability) as a function of minimum companion mass
for the model mass distributions. The upper-left panel corresponds to
the distributions with Ida \& Lin planetary masses while the
upper-right panel corresponds to Mordasini planetary masses. The
bottom panels compare scale factors as a function of mass between the
Ida \& Lin and Mordasini distributions with (bottom-left)
and without (bottom-right) brown dwarfs.}
\end{figure*}

\begin{table}
\begin{center}
\scriptsize
\caption{\label{tab:realScales} Scale Factors for Model Mass Distributions with Brown Dwarf Companions}
\begin{tabular}{lccccc}
\tableline
\multirow{2}{*}{Regime} & \multirow{2}{*}{Mass Range} & \multicolumn{2}{c}{Ida \& Lin\footnote[1]{Mass distribution using Ida \& Lin companion masses.}} & \multicolumn{2}{c}{Mordasini\footnote[2]{Mass distribution using Mordasini companion masses.}}\\\cline{3-6}
& & $\alpha$ & $\langle f\rangle$ & $\alpha$ & $\langle f\rangle$\\
\tableline
Earths/Super-Earths & $0.1M_\earth - 10M_\earth$ & -1.5 &1.12 & -1.4 & 1.14\\
Neptunes & $10M_\earth - 100M_\earth$ & -1.4 & 1.07& -1.7&1.19\\
Jupiters & $100 M_\earth - 10^3M_\earth$ & -0.9& 0.99 & -1.1 &0.99\\
Super-Jupiters & $3M_{\rm Jup} - 13M_{\rm Jup}$ & -2.2 & 1.42& -1.1 &1.12\\
Brown Dwarfs & $13M_{\rm Jup} - 0.07M_\sun$ & -0.9 &0.61 &-1.5 &0.86\\
Stars & $0.07M_\sun - 1M_\sun$ & -0.2 & 1.80&-0.1&1.79\\
\tableline
\end{tabular}
\end{center}
\end{table}
We plot the scale factor against mass in Figure \ref{fig:Ptformasses}.
For most mass ranges, the scale factor $f$ for a given $M_c\sin i $
differs little between the different mass distributions, with the most
pronounced discrepancy happening in the brown dwarf regime below the
deuterium-burning limit.  In Table \ref{tab:realScales}, we list the average scale factor values in each mass regime for the two mass distributions with brown dwarfs. There are several important points to note
about the distribution of scale factors.
\begin{itemize}
\item In the Jupiter mass regime
($100\textrm{M}_\earth-10^{3}\textrm{M}_\earth$), the slope of the
mass distribution is reasonably well-approximated by a power-law with index
$\alpha \approx -1$, so $f_{-1} = 1$ and the prior and posterior
transit probabilities are very similar. Consequently, using the
posterior probability in this mass regime, rather than the prior,
would not substantially change target selection criteria for transit
follow-up observations (assuming the prior on the mass function in
this regime is realistic).
\item In the Super-Jupiter regime ($3\textrm{M}_{\rm Jup} - 13
\textrm{M}_{\rm Jup}$), the transit probabilities are boosted by $\sim
20-50\%$.
\item The posterior transit probabilities for Neptunes and
Super-Earths are also boosted relative to the prior transit
probabilities.
\item The posterior transit probabilities for companions with minimum
mass near the high-mass end of the brown dwarf desert have very low
posterior transit probabilities and are thus poor targets for transit
follow-up searches.  These systems are simply much more likely to be
low-mass stars seen at relatively low inclination angle $i$.
\end{itemize}
Together, these results suggest that RV detected planets with $M_c\sin
i$ in the mass range of Earths, Super-Earths, Neptunes, or
Super-Jupiters are generally better-than-expected targets for transit
follow-up, whereas companions in the BD desert are poor targets for
transit follow-up.

\section{Application to Known RV Systems: Identifying Promising Systems for Photometric Follow-Up}\label{sec:RVs}

In this section, we apply the posterior transit probabilities derived
from our model mass distributions to known RV planets in order to
identify specific systems that might be significantly better
candidates for transit follow-up observations than would be expected
based on the naive prior transit probability.

Before doing so, however, it is worth asking whether or
not there is any evidence that the planet mass distributions we have
adopted have any correspondence to the true mass distribution of
planetary companions (c.f. \citealt{Mordasini2009}).  

Estimates of the true mass distribution of Jupiter-mass planets
discovered by RV surveys generally find distributions that are roughly
flat in $\log M_c$ (e.g. \citealt{Watson2010}), consistent with mass
functions used here and suggesting that the posterior and prior
transit probabilities for such companions should be similar.  It is
worth noting that the first transiting planet, HD 209458 b \citep{Charbonneau2000,Henry2000}, was
found ``right on time'' based on adopting the prior transit probability for
hot Jupiters of $\sim 10\%$, and given the $\sim$10 hot Jupiter systems that
were known at the time.

It is also interesting to note that there are two transiting
Super-Jupiters and three transiting Jupiters among the planets
orbiting solar-type stars that were originally discovered by RV,
despite the fact that super-Jupiters are much rarer than Jupiter-mass
companions among short-period systems.  This is consistent with the prediction
of the mass functions adopted here that the posterior transit probability
of Super-Jupiters is higher than the posterior transit probability of Jupiters.  

While this anecdotal evidence based on the rates of transiting Jupiters
and Super-Jupiters is intriguing, there are of course other factors
that strongly affect the transit probability, including the semimajor
axis, argument of periastron, and host radius.  To account for these
effects, we compared the number of RV-detected planets that have been
identified to transit to the expected number of transiting planets
based on the naive prior transit probability.  We took our exoplanet
sample from the Exoplanet Orbit Database on 19 April 2013
\citep{Wright2011}. We included only the 380 RV planets in the
database whose eccentricity $e$, semimajor axis $a$, angle of
periastron $\omega$ and host star's radius $R_{*}$ were listed. We
also included the eight planets first detected by RV and
subsequently found to transit. For each RV planet, we found the prior
transit probability from its orbital parameters and using the general
form of the prior transit probability given in Equation
\eqref{eq:priorP}, assuming $R_{*} \gg R_{c}$.  We then partitioned the
planets into three 1-dex bins of $\log(M_c\sin i)$.  We estimated the
expected number of transits in each bin by adding the prior transit
probabilities of the planets in that bin.  We divided the actual
number of transiting planets in each bin by the predicted number of
transiting planets.

\begin{figure*}
\plotone{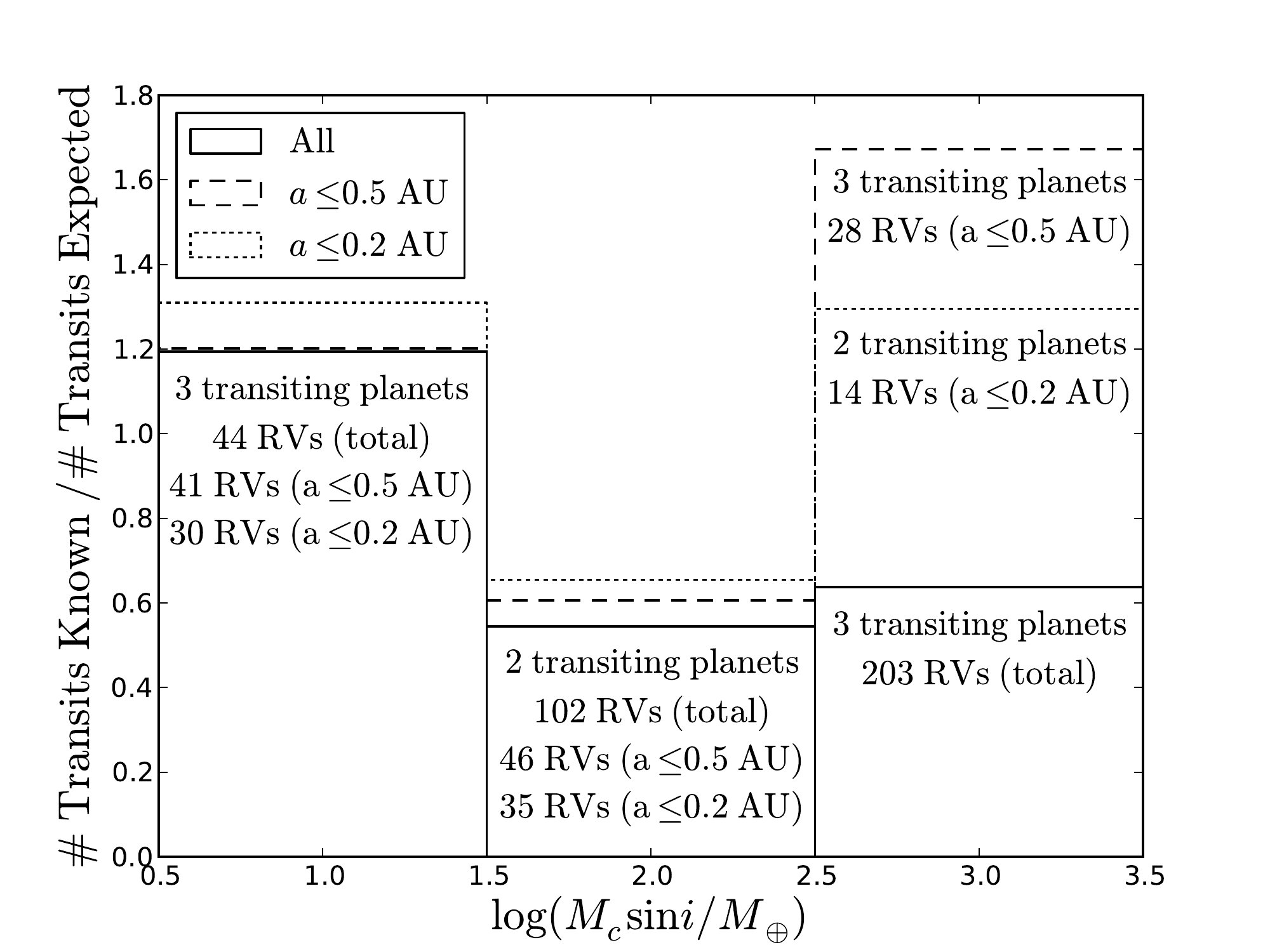}
\caption{\label{fig:relativentrans} The ratio of transits to the
expected number of transits as a function of $M_c\sin i$, plotted for
all planets in the sample (solid line), planets with semimajor axes $a
\leq 0.5$AU (dashed line) and planets with semimajor axes $a \leq
0.2$AU (dotted line).}
\end{figure*}

This ratio of the number of known transiting planets to the number
predicted based on the prior transit probability is shown in Figure
\ref{fig:relativentrans}.  Note that the ratio plotted is strictly a
lower limit to the actual ratio of transiting planets to predicted
number of transiting planets, since there may exist transiting planets
among the sample of RV detected systems that have not yet been
identified to transit.  In particular, identifying or definitively
excluding transits of long period systems is quite difficult because
of the infrequent transit opportunities and generally large
uncertainties in the predicted times of inferior conjunction.  With
this in mind, we also show the ratio after restricting to RV systems
with semiamajor axes $a\le 0.5$ AU and $a\le 0.2$ AU.  Presumably, the
systems in these subsets have been more thoroughly vetted for
transits, and thus the sample of transiting planets is more nearly
complete.
Because of the small number of known transiting systems and the likely
incompleteness, we caution against drawing any strong conclusions.
Nevertheless, for the bins in which there are a non-zero number of
known transiting systems, the results are generally consistent with
the expectations from the mass functions adopted here.  In particular,
the observed number of transiting planets
with minimum mass between $M_{\rm Jup}$ and $10M_{\rm Jup}$ ($M_c \sin i = 10^{2.5}\textrm{M}_\earth-10^{3.5}
\textrm{M}_\earth$) transiting planets is apparently higher than one would
expect based on the naive prior transit probability.

\subsection{Posterior-Prior Comparison and Transit Candidate Identification}\label{subsec:priorpost}
\begin{figure*}
\plotone{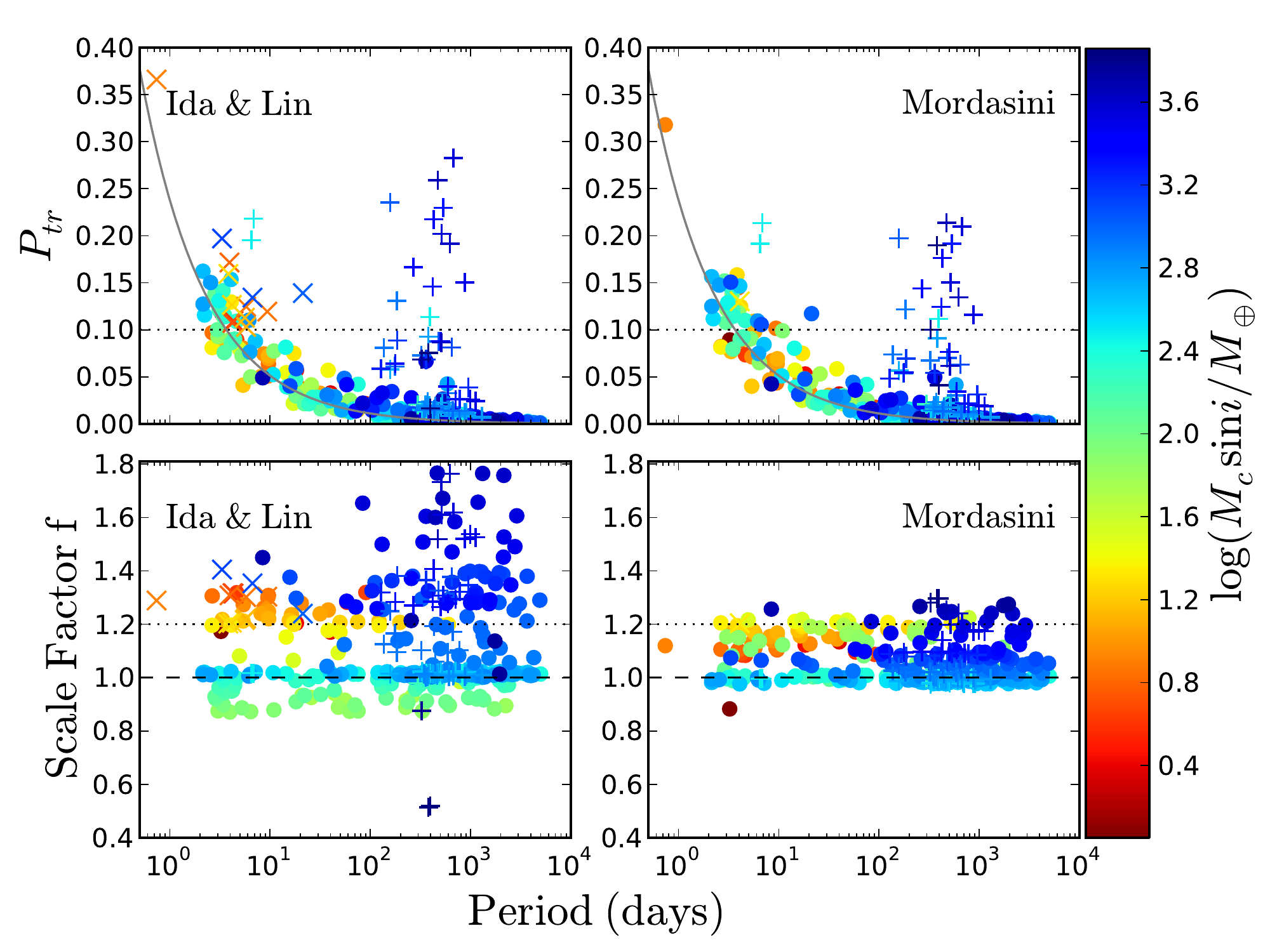}
\caption{\label{fig:planperprob} \emph{Top panels}: Posterior
probability as determined using the Ida \& Lin mass
distribution (left) and Mordasini distribution (right) for
RV-detected planets that have a full list of orbital parameters in the
Exoplanets.org Database.  The solid grey line shows
the simple prior transit probability for circular orbits, a
star with a solar density, and a planet with a small mass and
radius compared to the star.  The dotted line marks posterior transit
probabilities above 10\%. \emph{Bottom panels}: Scale factors for the
same planets. The dashed line denotes a posterior probability equal to
the prior, while the dotted line denotes a posterior that has a 20\%
boost relative to the prior. In all panels, bluer colors mark more
massive planets, a plus sign denotes planets whose host stars have $R_*
\geq 2R_\sun$ and X marks planets that have $P_{tr} \geq 0.1$ and
$f_\alpha \geq 1.2$; the latter are listed in Table
\ref{tab:candidates}}.
\end{figure*}

To compare the prior and posterior transit probabilities for known RV
planets, in Figure \ref{fig:planperprob} we plot the posterior probability
versus $T$, following Kane \& Von Braun (2008), as well as the ratio of the
posterior and prior transit probabilities.   The prior transit
probabilities were estimated with Equation \ref{eq:priorP} assuming
$R_p \ll R_*$, the values of $e$, $\omega$, $R_*$, $M_*$, and $a$ as given by the
Exoplanets Orbit Database.\footnote{The semimajor axes in this
database were derived using Kepler's
Third Law and assuming that the true companion mass is equal to its
minimum mass \citep{Butler2006,Wright2011}} We determine posterior
transit probability scalings using both the Ida \& Lin and
Mordasini model mass distributions that include brown dwarfs.
To find the posterior transit probability scaling for a given RV
system, we simply choose the value of the posterior transit
probability scaling for the 0.25 dex bin in $\log(M_c \sin i)$ in which the RV
companion is located in Figure \ref{fig:massdist}.

The overall trend of the transit probabilities versus $T$ is similar
for all cases, and simply reflects the scaling $P_{tr} \propto X
\propto T^{-2/3}$.  To guide the eye, in Figure \ref{fig:planperprob}
we plot the simple prior transit probability for circular orbits, a
star with a solar density, and a planet with a small mass and
radius compared to the star: $P_{tr,0}=(3\pi/G)^{1/3}\rho_\odot^{-1/3}T^{-2/3}$.  Deviations
from this fiducial probability are due to eccentricity, orientation of
the orbit ($\omega$), variations in $\rho_*$, and finally, in the case of
the posterior probabilities, variations in the posterior scaling
factor.  The vertical grouping of systems with large periods ($10^2
\text{days} \lesssim T \lesssim 10^3 \text{days}$) and relatively
large transit probabilities are planets orbiting giant stars with
large radii. Although the transit probabilities are large for these
systems, any transits in such systems would be very shallow and very
long.  Specialized methods are generally needed to detect such
transits; these are discussed in detail in \citet{Assef2009}.

The transit scale factor (ratio of posterior to prior transit
probabilities) inferred from the Ida \& Lin model shows considerably more
variance than that inferred from the Mordasini model.  This simply
reflects fact that the Mordasini mass distribution is smoother than
that of the Ida \& Lin distribution in the planetary regime (see
Fig.~\ref{fig:massdist}).  Also, because the Mordasini distribution of
companion masses increases nearly monotonically toward lower mass down
to $\sim M_\oplus$, the scale factors inferred from this model are
never substantially less than unity for the known exoplanets.  On the
other hand, the Ida \& Lin mass distribution shows a clear local minimum
near $\sim 30M_\oplus$ (a consequence of the ``planet desert'' 
discussed in \citealt{IdaLin2004}), and
therefore the inferred scale factors for planets with minimum mass
somewhat larger than this minimum are significantly less than unity.
For the Ida \& Lin model, we find that the posterior probability is
less than the prior probability for $\sim 19\%$ the planets in the
sample, as opposed to $\sim 34\%$ for the Mordasini distribution.

More interesting, however, is the fact that we infer a posterior
probability that is markedly higher for some planets, particularly for
the Ida \& Lin model.  These are the ones we will focus on here.  To
identify such promising transit candidates, we begin by eliminating
giant star hosts with $R*> 2.5 R_\sun$ (marked with plus signs in Figure
\ref{fig:planperprob}) from consideration.  From the remaining
planets, we choose planets from either model that have both a high
posterior transit probability of $P_{tr} \geq 0.1$, and a high prior
transit scale factor of $f \ge 1.2$, i.e. for which $P_{tr} >
1.2P_{tr,0}$.  These planets lie above the dotted lines in both the
top and bottom panels of Figure \ref{fig:planperprob}, and are marked
with an X.

These cuts leave us with a sample of fourteen planets, which are
listed in Table \ref{tab:candidates}.  Thirteen planets pass the cuts
from the Ida \& Lin distribution, whereas only HD 47186 b passes from
the Mordasini distribution.  None of the candidates pass the cuts
from both distributions.  Among these fourteen, 55 Cnc e and HD 17156
b have already been shown to transit, whereas transits in HD 40306 b,
HIP 14810 b, and $\tau$ Boo b have been conclusively ruled out.  The
sum of the posterior transit probabilities of the remaining candidates
is $\sim$1.15, suggesting that one transiting system may lurk amongst
these systems.  Detecting or excluding transits from most of these systems
will be challenging from the ground, as most have predicted depths
of $\sim$$0.05-0.1\%$, based on the minimum mass and a mass/radius relation
of $R_c= R_\oplus (M_c/M_\oplus)^{0.53}$ \citep{weiss2013}.  
\begin{table*}
\begin{center}
\caption{\label{tab:candidates} Transiting Planet Candidates}
\begin{tabular}{lccccccccc}
\tableline
Name & $M_c \sin i (\textrm{M}_{\rm Jup})$ & Period (days) & $M_* (\textrm{M}_\sun)$ & $R_* (R_\sun)$ & V (mag)  & Transit Depth\footnote[1]{Transit depth $(R_c/R_*)^2$,
assuming the minimum mass and a mass-radius relation $R_c= R_\oplus (M_c/M_\oplus)^{0.53}$ \citep{weiss2013}.}  & $P_{tr}$ & $P_{tr,0}$ & $f_\alpha$\\
\tableline
61 Vir b & 0.016 & 4.215 & 0.942 & 0.980 & 4.9 & $4.92 \times 10^{-4}$& 0.107 & 0.081 & 1.317 \\
BD -08 2823 b & 0.046 & 5.600 & 0.740 & 1.279 & 10 & $8.82 \times 10^{-4}$& 0.113 & 0.093 & 1.216 \\
HD 10180 c & 0.042 & 5.760 & 1.060 & 1.109 & 7.3 & $1.05 \times 10^{-3}$& 0.103 & 0.084 & 1.221 \\
HD 125612 c & 0.058 & 4.155 & 0.902 & 1.043 & 9.0 & $1.70 \times 10^{-3}$& 0.126 & 0.105 & 1.206  \\
HD 1461 b\footnote[2]{\citet{Rivera2010} did not definitively rule out transit.} & 0.024 & 5.773 & 1.026 & 1.130 & 6.6 & $5.67 \times 10^{-4}$ & 0.125 & 0.096 & 1.302 \\
HD 181433 b & 0.024 & 9.374 & 0.8 & 1.008 & 8.4 & $7.03 \times 10^{-4}$& 0.119 & 0.092 & 1.304 \\
HD 215497 b & 0.021 & 3.934 & 0.872 & 1.107 & 9.1 & $5.10 \times 10^{-4}$& 0.172 & 0.131 & 1.308 \\
HD 219828 b & 0.062 & 3.834 & 1.24 & 1.468 & 8.0 & $9.24 \times 10^{-4}$& 0.159 & 0.133 & 1.203\\ 
HD 47186 b\footnote[3]{Candidate using posterior from Mordasini mass distribution.} & 0.071 & 4.085 & 1.0 & 1.131 & 7.6 & $1.79 \times 10^{-3}$ & 0.130 & 0.108 & 1.204\\
55 Cnc e\footnote[4]{Known transiting planet.}\ & 0.026 & 0.737 & 0.905 & 0.943 & 6.0 & $4.54 \times 10^{-4}$& 0.366 & 0.284 & 1.289\\
HD 17156 b\footnotemark[4] & 3.30 & 21.217 & 1.285 & 1.507 & 8.2 & $5.29 \times 10^{-3}$& 0.139 & 0.112 & 1.240\\
HD 40307 b\footnote[5]{Transits excluded.} & 0.013 & 4.312 & 0.740 & 0.839 & 7.2 & $5.33\times 10^{-4}$ & 0.109 & 0.083 & 1.315\\
HIP 14810 b\footnotemark[5] & 3.87 & 6.674 & 0.990 & 1.320 & 8.5 & $9.11\times 10^{-2}$ & 0.134 & 0.099 & 1.353\\
$\tau$ Boo b\footnotemark[5] & 4.17 & 3.312 & 1.341 & 1.418 & 4.5 & $8.52 \times 10^{-2}$ & 0.197 & 0.140 & 1.405\\
\tableline
\end{tabular}
\end{center}
\end{table*}

\section{Additional Issues Affecting the Transit Probability}\label{sec:issues}

We now turn our attention to two issues that, to our
knowledge, have not been previously discussed and can, in principle,
subtly affect transit probability estimates.  First, we consider the
uncertainty in the transit probability arising from uncertainties in
the input parameters that are used to estimate the transit
probability.  Second, we consider the effect of the dependence of the
semimajor axis (and so the transit probability) on the orbital
inclination.  In both instances, we find that the quantitative changes
to the estimated transit probability are likely to be negligible in
most instances, although they can be larger and therefore more
important in some special cases.

We begin by writing down the expression for the posterior transit probability
in the limit of small transit probabilities (i.e., $X \ll 1$), in which case
we can write $P_{tr}$ as a constant factor $f$ times the prior transit 
probability,\footnote{Formally, we only demonstrated this to be true for power-law
priors for $M_c$, but we expect it to be generally true for any prior that is smooth
at $M_c=M_0$.}
\begin{equation}
P_{tr} \simeq f X = f \frac{R_*+R_c}{a} g(e,\omega).
\label{eqn:ptrpost}
\end{equation}
where we have defined 
\begin{equation}
g(e,\omega) \equiv \frac{1+e \sin \omega}{1-e^2}.
\label{eqn:geomega}
\end{equation}

It is instructive to deconstruct this expression for the transit
probability in terms of the quantities that can be measured directly
from the RV data and those that must be inferred or assumed from
external information.  The relevant RV observables are $T, e,$ and
$\omega$.  Using Newton's form of Kepler's Third law and defining the
bulk density of the star $\rho_* \equiv 3M_*/(4\pi R_*^3)$, the
transit probality can be written as
\begin{equation}
P_{tr} = \left(\frac{3\pi}{G}\right)^{1/3} f \rho_*^{-1/3} (1+q)^{-1/3} (1+r) T^{-2/3} g(e,\omega),
\label{eq:ptrdensity}
\end{equation}
where we have defined the mass ratio $q\equiv M_c/M_*$ and the radius
ratio $r\equiv R_c/R_*$.  Thus, in the limit that $r\ll 1$ and $q\ll 1$ the only
parameter of the star that enters into the transit probablity is the
density $\rho_*$.  In fact, one can write,
\begin{equation}
P_{tr} = (3\pi)^{1/3} f  (1+q)^{-1/3} (1+r) \left(\frac{t_{dyn}}{T}\right)^{-2/3} g(e,\omega),
\label{eqn:ptrtdyn}
\end{equation}
suggesting that the more fundamental physical quantity is, in fact, the
free-fall or dynamical time of the star $t_{dyn}\equiv
(G\rho_*)^{-1/2}$.

It is useful to examine and classify the parameters that enter into
Equation \eqref{eq:ptrdensity}.  The RV observables $T, e,$ and
$\omega$ will have uncertainties associated with them, and these
uncertainties may be correlated.  The stellar density $\rho_*$ must be
inferred from external information, i.e.\ from the stellar
temperature, surface gravity, and metallicity derived from a
high-resolution spectrum (perhaps combined with a parallax
measurement) or directly from astroseismology.  Both the RV parameters
and stellar densities will have measurement uncertainties associated
with them. We discuss the effect of these uncertainties in Section
\ref{subsec:Errors}.  Finally, the posterior scaling factor $f$ and
the radius ratio $r$ must simply be assumed.  These may be uncertain,
but these are not statistical uncertainties in the traditional sense;
rather, they are systematic uncertainties associated with the prior.
We will not consider these further.  The last parameter, $q$, arises
from the fact that the period $T$ is observed but it is the semimajor
axis that determines the transit probability.  In fact, the companion
mass $M_c$ has a definite value at the inclination at which transits
occur, and so this can be determined.  We explore this latter issue in Section
\ref{subsec:a}.

\subsection{Measurement Uncertainties}\label{subsec:Errors}
Discussion of transit probabilities --- prior or otherwise --- is
often unaccompanied by any discussion of the uncertainties in the
measurements of the relevant orbital parameters. Consider Equation
\eqref{eq:ptrdensity}, which can be
rewritten as $\log P_{tr} = \log f + \log (1+r) -
\frac{1}{3} \log (1+q) - \frac{1}{3} \log \rho_* - \frac{2}{3} \log T
+ \log g(e,\sin \omega) + const$.  Again,
we will ignore the uncertainty in
$f$ and $r$ here, and we discuss the effect of $q$ in the next
section.  The uncertainty in the transit probability $\sigma_{P_{tr}}$
due to uncertainties in the remaining quantitities is
\begin{equation}
\label{eq:errorprop}
\left(\frac{\sigma_{P_{tr,0}}}{P_{tr,0}}\right)^2 = \frac{1}{9} \left(\frac{\sigma_{\rho_*}}{\rho_*}\right)^2 + \frac{4}{9}\left(\frac{\sigma_T}{T}\right)^2 + \left(\frac{\sigma_{g(e,\sin \omega)}}{g(e,\sin \omega)}\right)^2,
\end{equation}
where we have assumed that the uncertainties in $\rho_*$, $T$, and
$(e,\omega)$ are uncorrelated, which is generally a good assumption.
We note that uncertainty in the transit probability depends fairly
weakly on the the stellar density $\rho_*$.  Furthermore, the period
$T$ is usually comparatively well-measured, so despite the higher
intrinsic sensitivity of $P_{tr}$ to it, it generally contributes
negligibly to the total uncertainty.

\begin{figure*}
\plotone{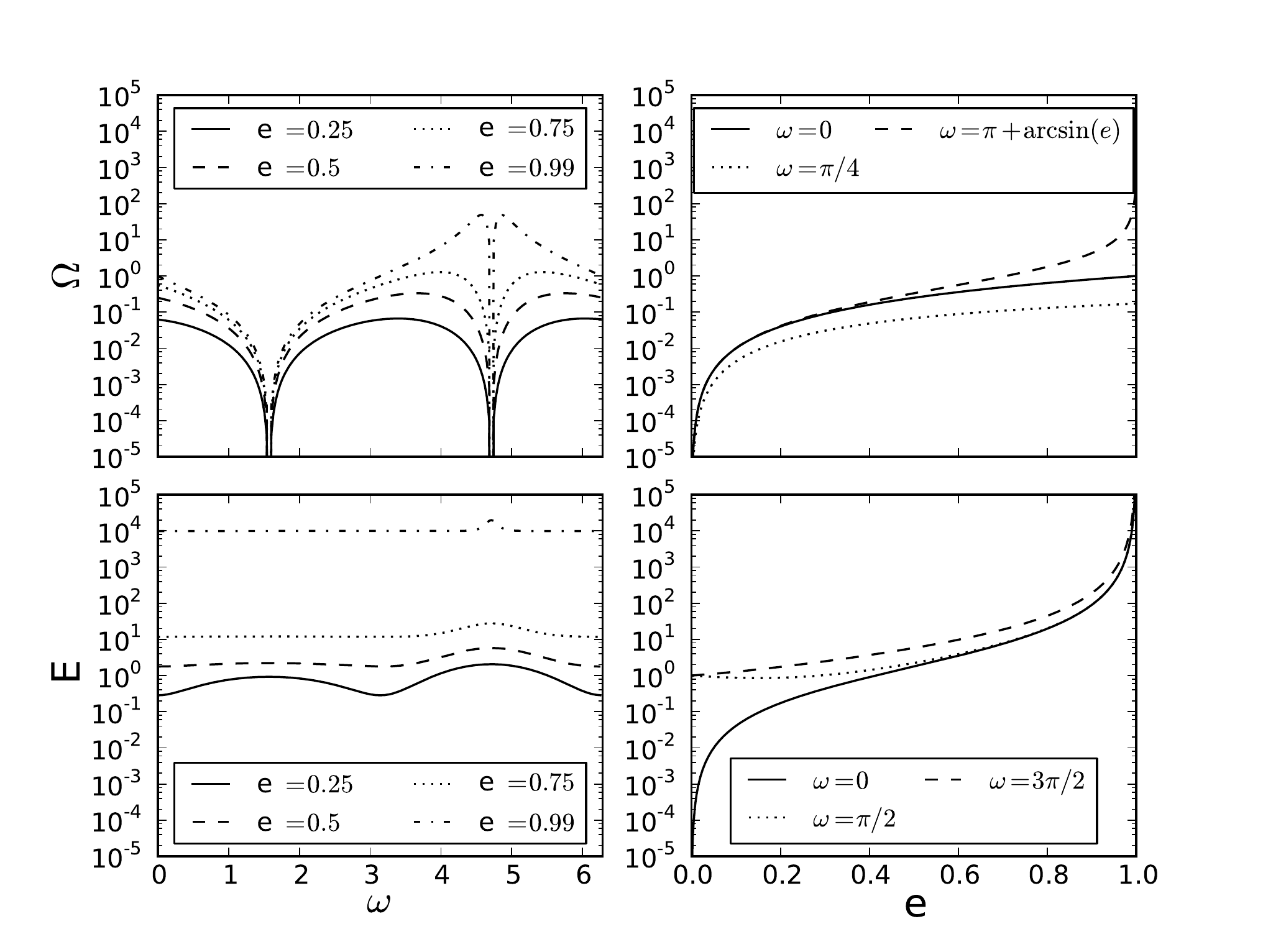}
\caption{\label{fig:ew} Contributions to the transit probability error
from $\Omega$ (top) and $E$ (bottom) as functions of $\omega$ (left)
and e (right) in the uncorrelated limit.}
\end{figure*}

Turning to $e$ and $\omega$, we note that the expression for
$\sigma_{g(e,\sin \omega)}$ depends on the covariance of $e$ and
$\omega$.  For simplicity, we consider only the limit in which these
parameters are uncorrelated:
\begin{equation}
\label{eq:ewuncorr}
\left(\frac{\sigma_{g(e,\sin \omega)}}{g(e,\sin \omega)}\right)^2 \approx 
E \sigma^2_e +\Omega \sigma^2_\omega,
\end{equation}
where we have defined 
\begin{equation}
E \equiv \frac{4e^2}{(1-e^2)^2} + \frac{\sin^2\omega}{(1+e\sin
\omega)^2}
\label{eqn:defE}
\end{equation}
and 
\begin{equation}
\Omega \equiv \left[\frac{e\cos\omega}{1+e\sin\omega}\right]^2
\label{eqn:defOmega}.
\end{equation}
Figure \ref{fig:ew}
shows the prefactors $E$ and $\Omega$ as a function of $e$ and
$\omega$.

It is instructive to consider the behavior of $E$ and $\Omega$ is the
limit of $e \ll 1$.  To first order in $e$, we have that $E \simeq
\sin^2\omega - 2e \sin^3 \omega$.  For $\omega=0$ and $\omega=\pi$,
i.e. orbits such that the semimajor axis is perpendicular to the line
of sight, the contribution of the eccentricity to the uncertainty in
$P_{tr}$ vanishes.  In addition, for $e=0$, there is a finite
contribution to the uncertainty in $P_{tr}$ for all values of $\omega$
except $0$ and $\pi$, reaching a maximum of $E=1$ for $\omega=\pi/2$
or $3\pi/2$; these values correspond to orbits such that the semimajor
axis is along the line of sight.  With regards to $\Omega$, there is
no contribution to the uncertainty in $P_{tr}$ to first order in $e$,
and we can therefore expect the contribution of the uncertainty in $e$
to dominate over that of $\omega$.  We conclude that for small $e$,
uncertainties in $e$ and $\omega$ generally do not lead to large
uncertainties in $P_{tr}$, with the contribution to the fractional
uncertainty in $P_{tr}$ at most of order $\sigma_e$.

On the other hand, for orbits with large eccentricity, the
uncertainties in $e$ and $\omega$ can have have a substantial effect
on the uncertainty in the transit probability.  First consider the
full expression for $E$.  It is clear that since both terms are
positive and the first term in $E$ does not depend on $\omega$, $E$ is
minimized at $\omega=0$ for all $e$.  Furthermore, $E$ diverges as
$e\rightarrow 1$ for all $\omega$.  Next, consider the full expression
for $\Omega$.  The contribution of $\omega$ to the uncertainty
vanishes for $\omega = \pi/2$ or $3\pi/2$, regardless of eccentricity.
$\Omega$ is maximized at $\omega=\pi + \arcsin{e}$ and $\omega=2\pi -
\arcsin{e}$, with maximum value of $\Omega_{max}=e^2/(1-e^2)$ that
diverges as $e\rightarrow 1$.

\subsection{True versus Minimum Semimajor Axis}\label{subsec:a}

In order to estimate the transit probability, one needs an estimate of
the true semimajor axis.  However, one measures only the period $T$
and the mass function $\mf$.  These, together with an estimate of
$M_*$, only allow one to measure a {\it minimum} semimajor axis $a_{min}$.
This minimum semimajor axis can be determined by first using the expression
for the mass function (Equation \ref{eqn:massfunc}) to determine the
minimum companion mass $M_{c,min}$ for $\sin i=1$, and then using this
to solve for $a$ using Newton's version of Kepler's Third Law
(Equation \ref{eqn:semi}).  Since $a_{min}$ is the minimum semimajor axis, the
transit probability estimated in this way will be an
overestimate of the true transit probability. 

Since the expression for $a_{min}$ is complicated, we will consider a somewhat
simpler estimate for $a$, 
\begin{equation}
a_0 \equiv \left(\frac{GM_*}{4\pi^2}\right)^{1/3} T^{2/3} = a(1+q)^{-1/3}.
\label{eqn:a0}
\end{equation}
The relation between the transit probability $P_{tr,a_0}$ estimated
using $a_0$ and the true transit probability is
\begin{equation}
P_{tr} = P_{tr,a_0} (1+q)^{-1/3} \simeq P_{tr,a_0}\left(1-\frac{1}{3}q\right),
\label{eqn:ptra0}
\end{equation}
where the rightmost expression holds for $q\ll 1$.  Thus, for small $q$, the fractional
amount by which one overestimates the transit probability using $a_0$ is $\sim q/3$.  
For a Jupter-mass companion, this is only $\sim 0.03\%$ --- essentially negligible.
For a companion at the hydrogen-burning limit, the error is $2.5\%$.  Of course, one
does not know the mass of the companion {\it a priori}.

Although the amount by which one overestimates the transit probability
is generally negligible, the true transit probability can nevertheless
be determined with no approximation.  This is accomplished by solving
for the value of $q$ at the critical inclination for a transit
$i_{min}$, and then using this value of $q$ to determine the true
semimajor axis.  We proceed by defining
\begin{equation}
\label{eq:Q1}
Q \equiv \frac{\mf}{M_*} = \frac{q}{(1+q)^{2/3}}(1-\cos^2 i)^{1/2},
\end{equation}
which we call the ``mass ratio function'' in analogy to the mass
function.  The true transit probability at the critical inclincation
is simply the cosine of the critical inclination, i.e., $P_{tr}= \cos
i_{min} = P_{tr,a_0}(1+q)^{-1/3}$.  We can then insert this expression
for $\cos i_{min}$ into Equation \eqref{eq:Q1} and, by defining $y
\equiv 1+q$, derive a polynomial equation in $y$ when $i=i_{min}$:
\begin{equation}
\label{eq:Q}
Q^2y^2 - (y-1)^2(y^{2/3} - P^2_{tr,a_0}) = 0.
\end{equation}
Recall that $Q$ and $P_{tr,a_0}$ are observables. 
This expression can be solved for $y$ with standard techniques.

\begin{figure*}
\plotone{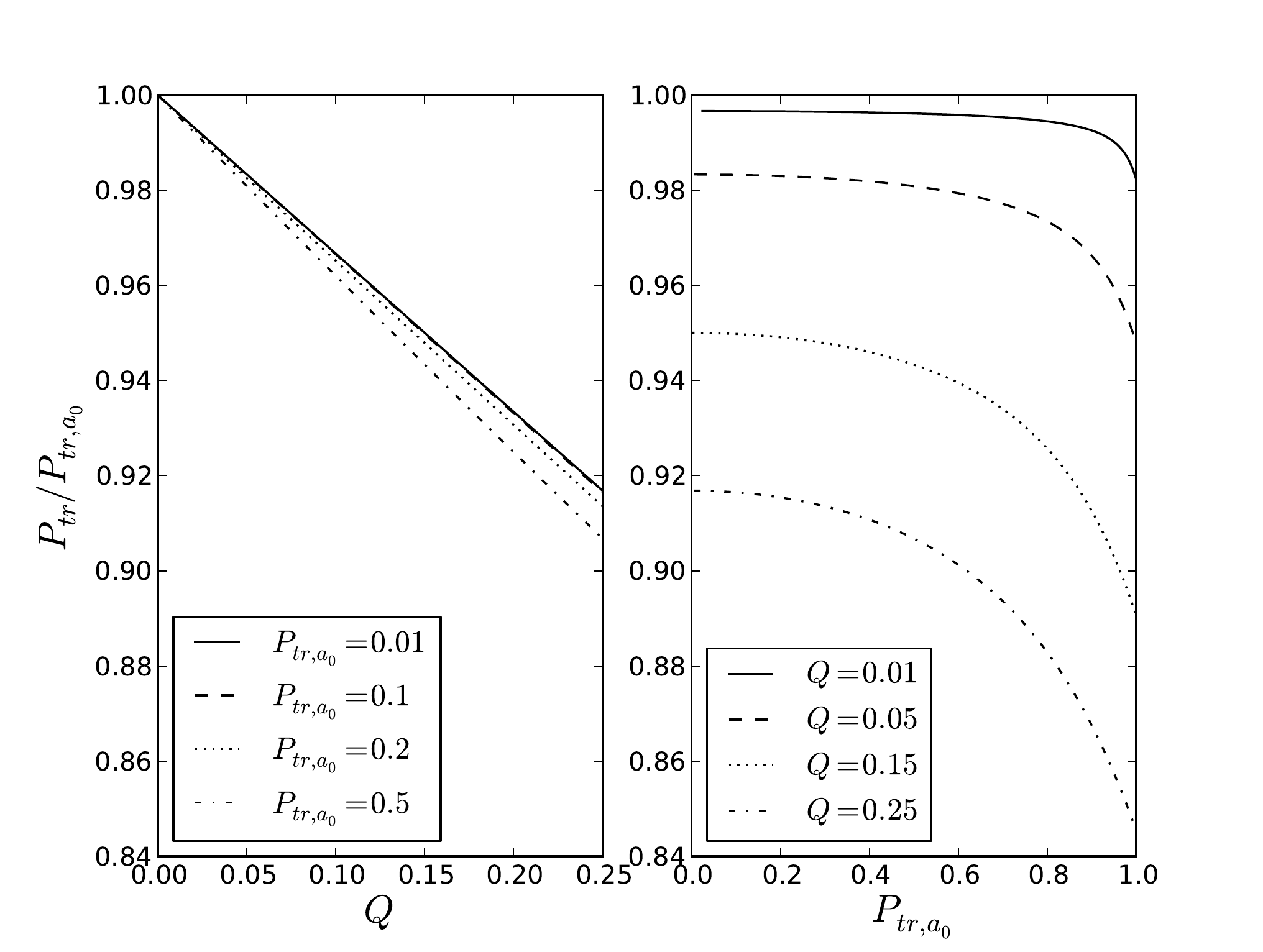}
\caption{\label{fig:QP} Comparison of the actual prior transit
probability, $P_{tr}$, and the prior transit probability calculated
with $a_0$, $P_{tr,a_0}$, plotted against the ``mass function
ratio'' $Q$ (left) and $P_{tr,a_0}$ (right). For reasonable values of $P_{tr,a_0}$ ($\lesssim 0.5$), using $a_0$ leads to overestimation of the prior by no more than $10\%$ over the entire range of $Q$.}
\hspace{300 mm}
\end{figure*}

Figure \ref{fig:QP} shows $P_{tr}/P_{tr,a_0} = y^{-1/3}$ as functions
of $Q$ and $P_{tr,a_0}$. For typical prior transit probabilities,
$P_0$ overestimates the actual prior transit probability by less than
$10\%$ over the entire range of $Q$.  Note that, in the limit that $q
\ll 1$,
\begin{equation}
\label{eq:Qqsmall}
q \simeq Q(1-P_{tr,a_0}^2)^{-1/2}.
\end{equation}
In the right panel of Figure \ref{fig:QP}, for $P_{tr,a_0} \ll 1$,
$P_{tr}/P_{tr,a_0} \simeq 1-Q/3$.

\section{Summary and Conclusion}\label{sec:conclusion}

We have derived a general expression for the \emph {posterior}
probability that a radial velocity companion with minimum mass $M_0=M_c \sin i$
also transits its parent star.  The posterior transit
probability depends on both the prior distribution of the orbital
inclination angle $i$ {\it and} the prior distribution of the true
companion mass $M_c$.  We evaluated this expression for power-law
distributions of $M_c$ and integer power-law indices $-3 \leq \alpha
\leq 3$, deriving exact analytic expressions for the posterior transit
probility for these cases.  We found that, for power-law distributions
in general, the posterior transit probability is well-approximated as a scalar
multiple $f_\alpha$ of the prior probability, with specific values
$f_{-3} = 3/2, f_{-2} = 4/\pi, f_{-1} = 1, f_{0} = 2/\pi, f_{1} =
1/\arctanh(\cos i_{min}), f_2 = \tan i_{min}$, and $f_3 =
2/[\arctanh(\cos i_{min}) + \cot (i_{min}) \csc (i_{min})]$, where
$i_{min}$ is the minimum inclination angle corresponding to the
maximum companaion mass such that $M_{c,max}/M_0 = 1/\sin i_{min}$.
For $\alpha \ll 0$, $f_\alpha \sim (-2\alpha/\pi)^{1/2}$.

We then applied our findings to four synthetic but physically
motivated companion mass distributions. In all four cases, we find
that $\alpha\simeq -1$ for Jupiters
($100\textrm{M}_\earth-10^{3}\textrm{M}_\earth$), and therefore the
prior and posterior probabilities are very similar.  On the other
hand, we find $\alpha \simeq -1.5$ for Earths and Super-Earths
($0.1\textrm{M}_\earth-10\textrm{M}_\earth$), and $\alpha \simeq -2\ 
\textrm{to} -2.5$ for Super-Jupiters
($10^{3}\textrm{M}_\earth-13\textrm{M}_{\rm Jup}$). The posterior transit
probability for RV planets with masses in these regimes is therefore
boosted relative to the prior, so we may expect more transiting
planets in these mass regimes.  With transit surveys pushing well into
the Super-Earth regime, this result is encouraging.

For brown dwarfs ($13\textrm{M}_{\rm Jup}-0.07\textrm{M}_\sun$), the scale
factor of the posterior transit probability relative to the prior
depends in detail on the aridity of the brown dwarf desert, as well as
on the precise value of the minimum mass that is measuremed.  In
general, the scale factor drops as one moves from the Super-Jupiter up
through the brown dwarf mass regime.  In particular, objects with
minimum mass just above the driest part of the brown dwarf desert may
have strongly suppressed transit probabilities, since such an object
is more likely a stellar-mass companion being viewed pole-on than a
true brown dwarf.  For stellar companions,
($0.07\textrm{M}_\sun-1\textrm{M}_\sun$), $\alpha \approx 0$, and thus
the posterior transit probablity is generally smaller than the prior
probability in this regime.

Using these mass distributions and the corresponding scale factors, we
estimated posterior transit probabilities for RV-discovered planets in
the Exoplanets.org Database \citep{Wright2011}. Selecting planets for
which the posterior transit probability is $>10\%$ and $>20\%$
larger than the prior probability, excluding companions around giants,
known transiting planets, and planets for which transits have been
ruled out, we found nine particularly promising transiting planet
candidates: 61 Vir b, BD -08 2823 b, HD 10180c, HD 125612 c, HD 1461
b, HD 181433 b, HD 215497 b, HD 219828 b, and HD 47186 b.

Finally, we discuss two issues that can subtly affect the calculation
of either the posterior or the prior transit probability. We find that
uncertainties in the eccentricity $e$ and argument of periastron
$\omega$ generally do not yield significant uncertainties in the
transit probability, provided $e$ is small. However, measurement
uncertainties in $e$ and $\omega$ for companions in
highly-eccentric orbits can have a significant impact on the
transit probability, with the uncertainty in $e$ being amplified by a
factor that diverges as $e \to 1$ for all $\omega$, and the
uncertainty in $\omega$ being amplified by a factor that also diverges
with increasing $e$ for the specific case of $\omega = \pi +
\arcsin(e)$.  Additionally, we point out that the semimajor axis
typically used in determining the transit probability is typically
smaller than the true semimajor axis, which leads to an overestimate
of the true probability. We demonstrate that this overestimate is
negligible for systems with small transit probabilities, but becomes
more significant as the transit probability aproaches unity.

The primary difficulty with estimating the posterior transit
probablity is that the mass distribution of low-mass companions is
poorly known.  Nevertheless, even the theoretical distributions we
studied here likely provide a more accurate estimate than the naive
prior transit probability.  These distributions suggest that RV
companions with minimum masses in the Earth/Super-Earth and
Super-Jupiter regimes are promising targets for transit follow-up.
As measurements of the companion mass distribution become increasingly
robust, we look forward to more frequent use of the posterior transit
probability --- accompanied, of course, by an appropriate discussion
of assumptions, priors, and sources of uncertainty.

\acknowledgments\label{sec:acknowledgements}
We thank Shigeru Ida and Christoph Mordasini for allowing us to use results from
their planetary formation models in our posterior transit probability
calculations. We also thank Thomas Beatty, Christian Clanton, Calen
Henderson, Matthew Penny, and Tim Morton for their helpful discussions
and suggestions. This research has made use of the Exoplanet Orbit
Database and the Exoplanet Data Explorer at exoplanets.org.
\nocite{*}
\bibliographystyle{apj}
\bibliography{SG2013.bib}
\end{document}